\documentclass[%
aip,
cha,
amsmath,amssymb,
superscriptaddress,
reprint,
]{revtex4-1}
\usepackage{graphicx}          
\usepackage{subfigure}         
\usepackage{dcolumn}           
\usepackage{bm}                
\usepackage{bbm}               
\usepackage[mathscr]{eucal}    
\usepackage[mathlines]{lineno} 
\usepackage{mathtools}
\usepackage[utf8]{inputenc}
\usepackage[T1]{fontenc}
\usepackage{mathptmx}
\usepackage{etoolbox}
\usepackage{time}
\usepackage{physics}
\usepackage{hyperref}          
\hypersetup{
    pdfnewwindow=true,         
    colorlinks=true,           
    linkcolor=red,             
    citecolor=blue,            
    filecolor=green,           
    urlcolor=cyan              
}
\newcommand{\Schrodinger}{Schr{\"o}dinger}
\newcommand{\rmi}{{\rm i}}                     
\newcommand{\rme}{{\rm e}}                     
\newcommand{\bbC}{\mathbbm{C}}
\newcommand{\bbR}{\mathbbm{R}}
\makeatletter
\def\@email#1#2{%
 \endgroup
 \patchcmd{\titleblock@produce}
  {\frontmatter@RRAPformat}
  {\frontmatter@RRAPformat{\produce@RRAP{*#1\href{mailto:#2}{#2}}}\frontmatter@RRAPformat}
  {}{}
}%
\makeatother
\begin{document}
%
%
\preprint{LA-UR-30266}
\begin{flushright}\textbf{LA-UR-30266}\end{flushright}
\title{Exact trapped $N$-soliton solutions of the nonlinear \Schrodinger\ equation
using the inverse problem method}
\author{Fred~Cooper}
\email{cooper@santafe.edu}
\affiliation{
   Santa Fe Institute,
   1399 Hyde Park Road,
   Santa Fe, NM 87501, USA}
\affiliation{
   Center for Nonlinear Studies and Theoretical Division, 
   Los Alamos National Laboratory, 
   Los Alamos, NM 87545, USA}
\author{Avinash Khare}
\email{avinashkhare45@gmail.com}
\affiliation{
   Physics Department, 
   Savitribai Phule Pune University, 
   Pune 411007, India} 
\author{John F. Dawson}
\email{john.dawson@unh.edu}
\affiliation{
   Department of Physics,
   University of New Hampshire,
   Durham, NH 03824, USA}
\author{Efstathios G. Charalampidis}
\email{echarala@calpoly.edu}
\affiliation{
   Mathematics Department, 
   California Polytechnic State University, 
   San Luis Obispo, CA 93407-0403, USA} 
\author{Avadh Saxena} 
\email{avadh@lanl.gov} 
\affiliation{
   Center for Nonlinear Studies and Theoretical Division, 
   Los Alamos National Laboratory, 
   Los Alamos, NM 87545, USA}
\date{\today, \now \ PST}
\begin{abstract}
In this work, we show the application of the ``inverse problem'' method to construct exact $N$
trapped soliton-like solutions of the nonlinear \Schrodinger\ or Gross-Pitaevskii equation (NLSE
and GPE, respectively) in one, two, and three spatial dimensions. This method is capable of finding
the external (confining) potentials which render specific assumed waveforms exact solutions of the
NLSE for both attractive ($g<0$) and repulsive ($g>0$) self-interactions.  For both signs of $g$,
we discuss the stability with respect to self-similar deformations and translations. For $g<0$, a
critical mass $M_c$, or equivalently the number of particles, for instabilities to arise can often
be found analytically. On the other hand, for the case with $g>0$ corresponding to repulsive
self interactions which is often discussed in the atomic physics realm of Bose-Einstein condensates (BEC),
the bound solutions are found to be always stable. For $g<0$, we also determine the critical mass
numerically by using linear stability or Bogoliubov-de Gennes analysis, and compare these results
with our analytic estimates. Various analytic forms for the trapped $N$-soliton solutions are discussed,
including sums of Gaussians or higher-order eigenfunctions of the harmonic oscillator Hamiltonian.
\end{abstract}
\maketitle
%
%
\begin{quotation}
Understanding the behavior of trapped atoms in BECs requires the numerical study of
the existence, stability and spatio-temporal dynamics of solutions to the Gross-Pitaevskii
equation (GPE). Exact solutions of the GPE subject to external potentials offers a path
in which not only numerical simulations can be carried out for this purpose but analytical
estimates for the stability of coherent structures can be derived. In this work, we consider
the inverse problem method which is capable of determining suitable external potentials that
make specified $N$-trapped soliton wave functions exact solutions to the GPE. The stability
of these solutions is then studied using Derrick's theorem and energy landscape techniques.
We discuss potential realizations of trapped BECs in 1D, 2D, and 3D. Our theoretical results
on stability analysis are compared with spectral computations in the realm of Bogoliubov-de
Gennes analysis.
\end{quotation}
%
%
\section{\label{Intro}Introduction}

The nonlinear \Schrodinger\ equation (NLSE)~\cite{ablowitz-2004} has arguably been
the focal point of studies in nonlinear models because its ubiquitous envelope
equation arises in diverse physical contexts with a wide array of physical applications. Those include the description of the pulse propagation in nonlinear optical fibers~\cite{Hasegawa-1995,kivshar-2003}, the evolution of the envelope of modulated wave
groups~\cite{zakharov-1968,ablowitz-2011}, as well as the propagation of strongly dispersive
waves in plasmas~\cite{Kono-2010}, among many others. When the NLSE incorporates an external,
i.e., confining potential, it is often called the Gross-Pitaevskii equation (GPE) which is a
fundamental model for describing the static and dynamical properties of atomic Bose-Einstein
condensation (BEC) in the mean-field approximation~\cite{Gross-1961,Pitaevskii-1961,Pitaevskii-2015}.
Indeed, solutions (either obtained analytically or numerically) of the related GPE in multiple
well potentials are very useful in understanding the behavior of trapped atoms in BECs. Both
signs of the self-interaction coupling constant can be implemented when studying BECs, by varying
the external magnetic field near the Feshbach resonance~\cite{Cornish-2000}. Using such methods,
attractive self-interaction solitons have been found in BECs~\cite{Strecker-2002}.

There are various strategies for finding solutions to the NLSE for given external potentials.
Indeed, and for a given potential, one may linearize the NLSE (i.e., upon neglecting the nonlinearity
therein), and obtain an eigenvalue problem for the (discrete) energy levels (eigenvalues) and
quantum states (eigenfunctions) of the system. The resulting problem is of a Sturm-Liouville type,
i.e., a linear \Schrodinger\ equation, and may be solved either analytically~\cite{Landau-1989}
or numerically, see, e.g., Refs.~\onlinecite{PhysRevE.91.012924,PhysRevE.94.022207}. Its eigenvalues
coincide with the values of the so called chemical potential~\cite{Pitaevskii-2015} at which nonlinear
states bifurcate from. Then, for each eigenvalue (i.e., value of the chemical potential at the linear limit)
and respective linear state, one can continue the latter towards the nonlinear regime by varying the chemical
potential which itself controls the number of atoms in a BEC~\cite{Pitaevskii-2015}. This departure from the
linear limit is accomplished by using numerical continuation methods~\cite{Allgower-1990}.  Another strategy
for finding solutions to the NLSE revolves around starting with an approximate solution, and then varying the
potential to find a solution.

In the present article, we depart from these strategies, and use the so-called ``inverse problem''
method. Within this method, one chooses beforehand a wave function that has to be an exact solution of
the NLSE. This way, various external potentials can be constructed with an eye towards realizing them
experimentally. This method has previously been used by Malomed and Stepanyants~\cite{Malomed-2010} in
the standard NLSE to determine potentials that have exact Gaussian-like solutions. It has also been used
in Ref.~\onlinecite{cooper-2022} for potentials in the NLSE with arbitrary nonlinearity exponent. Recently,
the authors of the present work have shown how to find confining potentials in the NLSE which lead to constant
density, flat-top solitons in one, two, and three dimensions (denoted hereafter as 1D, 2D, and 3D, respectively)~\cite{cooper-2023}.
Herein, we consider wave function Ans\"atze corresponding to $N$-soliton pulses, and identify
the respective potentials that make them exact solutions to the NLSE in 1D, 2D, and 3D. Moreover, and since
the inverse problem method gives us exact solutions, we take this advantage in order to provide analytic estimates
for the critical mass for attractive self-interaction solitons above which the soliton becomes unstable.  Those are obtained
by using Derrick's theorem~\cite{Derrick-1964}, or by studying the energy landscape for translation deformations
of the soliton \cite{cooper-2022}.  We compare our analytical findings on stability and instability of the
soliton solutions against linear stability considerations by using the Bogoliubov-de Gennes\cite{Bogolyubov-1947,deGennes-1966}
(BdG) method.

The paper is structured as follows. In Sec.~\ref{s:STABILITY}, we present the main setup of the
inverse problem method together with the linear response equations. Multi-soliton solutions
in 1D, 2D, and 3D are discussed in Sec.~\ref{s:MultiSoliton} together with their response under
self-similar and translational deformations. In Sec.~\ref{s:LinearResponse} we study the linear
response equations and compare our findings against numerical simulations. Finally, we state our
conclusions in Sec.~\ref{s:conclusions}.

%
%
\section{\label{s:STABILITY}Inverse Problem Method for the confining potential and the linear stability of  the solutions}

We consider herein a collection of particles with mass $m = 1/2$, and contact interaction
strength $g$ which is described by a classical action. Upon confining the particles with
the introduction of an external potential denoted as $V(\vb{r})\in \bbR$, the nonlinear
\Schrodinger\ equation (NLSE) for this system~\cite{Pitaevskii-2015} is
then given by:
\begin{equation}\label{e:NLSE}
   \bigl \{\,
      -
      \laplacian
      +
      g \, |\psi(\vb{r},t)|^{2}
      +
      V(\vb{r}) \, 
   \bigr \} \, \psi(\vb{r},t)
   =
   \rmi \, \partial_t \psi(\vb{r},t) \>,
\end{equation}
where $\psi(\vb{r},t)$ is a complex-valued function, i.e., $\psi(\vb{r},t)\in\bbC$. Here we use units such that $\hbar=1$ (see, also Ref.~\onlinecite{cooper-2023}). It should be noted in passing that in the absence of the external potential (i.e., $V(\vb{r}) \equiv 0$), soliton solutions exist for both repulsive ($g>0$) interactions (see Ref.~\onlinecite{Gaidoukov-2021}), as well as attractive ($g<0$) interactions
(see Ref.~\onlinecite{10.1143/PTPS.55.284}).

Suppose that $u_0(\vb{r})\in\mathbb{R}$ is the solution to Eq.~\eqref{e:NLSE} at $t=0$. If we assume a time-dependent solution for $\psi(\vb{r},t)$ given by  the separation of variables ansatz:
\begin{equation}\label{e:psi-u}
   \psi(\vb{r},t)
   =
   u_0(\vb{r}) \, \rme^{-\rmi \, \omega t} \>,
\end{equation}
then Eq.~\eqref{e:NLSE} is written as:
\begin{equation}\label{e:ueq}
   \omega \, u_0(\vb{r})
   +
   \nabla^{2} u_0(\vb{r})
   -
   g \, u_0^2(\vb{r}) \, u(\vb{r})
   =
   V(\vb{r}) \, u(\vb{r}) \>.
\end{equation}
If we are considering the Gross-Pitaevskii equation (GPE)~\cite{Gross-1961,Pitaevskii-1961,Pitaevskii-2015} for BECs as a particular NLSE, then $\omega \rightarrow \mu_0$, where  $\mu_0$ is the chemical potential. (The connection between the NLSE and GPE is discussed among other places in Ref.~\onlinecite{cooper-2023}.) 

The potential that will make $\psi(\vb{r},t)=u_0(\vb{r}) \rme^{- \rmi \mu_0 t}$ an exact solution of the GPE is given by the (inverse) relation:
\begin{equation}\label{e:Vdef}
   V(\vb{r})
   =
   \mu_0 - g u_0^2(\vb{r}) + \frac{\laplacian u_0(\vb{r})}{u_0(\vb{r})} \>.
\end{equation}
It is therefore the task of the experimenter to create such a potential. It is important to now regard the potential $V(\vb{r})$ as so constructed to be \emph{external}, and is not varied with respect to $u_0(\vb{r})$. Since the potential is now fixed, the conserved energy is given by
\begin{equation}\label{e:ConsE}
   E_0
   =
   \int \dd[3]{x}
   \Bigl \{\,
      [ \grad{u_0(\vb{r})} ]^2
      +
      \frac{g}{2} \, u_0^4(\vb{r})
      +
      V(\vb{r}) \, u_0^2(\vb{r}) \,
   \Bigr \} \>,
\end{equation}
and the conserved norm which is related to the number of atoms in the BEC
(see, Ref.~\onlinecite{cooper-2023}) is given by
\begin{equation}\label{eq:cons_mass}
   M = \int \dd[3]{x} |\psi(\vb{r},t)|^2 \>.
\end{equation}
Soliton wave functions in 1D, 2D, and 3D are discussed in Sec.~\ref{s:MultiSoliton} below.

The linear stability of such solutions in the constructed potential is found by expanding the solution $\psi(\vb{r},t)$ in the form of power series in
$\varepsilon (\ll 1)$:
\begin{align}
   \psi(\vb{r},t) 
   &= 
   \psi_0(\vb{r},t) + \varepsilon \, \phi(\vb{r},t) + \dotsb
   \label{e:psiexpand} \\
   &=
   \rme^{- \rmi \mu_0 t} \, u_0(\vb{r})
   + 
   \varepsilon \, \phi(\vb{r},t) + \dotsb \,, 
   \notag
\end{align}
where $\mu_0$ is the chemical potential, and $u_0(\vb{r})$ is a particular
solution of the time-independent GPE~\cite{Gross-1961,Pitaevskii-1961,Pitaevskii-2015}.
To first order in $\varepsilon$, $\phi(\vb{r},t)$ and $\phi^{\ast}(\vb{r},t)$
satisfy:
\begin{align}
   &\begin{pmatrix}
     [\, h(\vb{r})  + g u_0^2(\vb{r}) \,] & g u_0^2(\vb{r})  \\
     - g u_0^2(\vb{r})  & - [\, h(\vb{r})  + g u_0^2(\vb{r}) \,]
   \end{pmatrix}
   \begin{pmatrix}
      \phi(\vb{r},t) \\
      \phi^{\ast}(\vb{r},t)
   \end{pmatrix}
   \notag \\
   &\hspace{3em}
   =
   \rmi \, \partial_t
   \begin{pmatrix}
      \phi(\vb{r},t) \\
      \phi^{\ast}(\vb{r},t)
   \end{pmatrix}  \>, 
   \label{e:DiffEq}
\end{align}
where $h(\vb{r})$ is the Hermitian operator:
\begin{align}
   h(\vb{r})
   &= -\laplacian + V_0(\vb{r}) \>,
   \label{e:hdef} \\
   V_0(\vb{r})
   &=
   V(\vb{r}) + g u_0^2(\vb{r})
   =
   \mu_0 + \frac{\laplacian u_0(\vb{r})}{u_0(\vb{r})}\>.
   \label{e:V0deff}
\end{align}
%
%
Solutions to the linear response equations~\eqref{e:DiffEq} are
discussed in Sec.~\ref{s:LinearResponse} below.

%
%
\section{\label{s:MultiSoliton}Multi-soliton solutions}
%
%
\subsection{\label{ss:1D}One dimension}

\begin{figure}[t]
   \centering
   \subfigure[\ $\rho_0(x)$]
   { \label{f:fig1a}
     \includegraphics[width=0.9\linewidth]{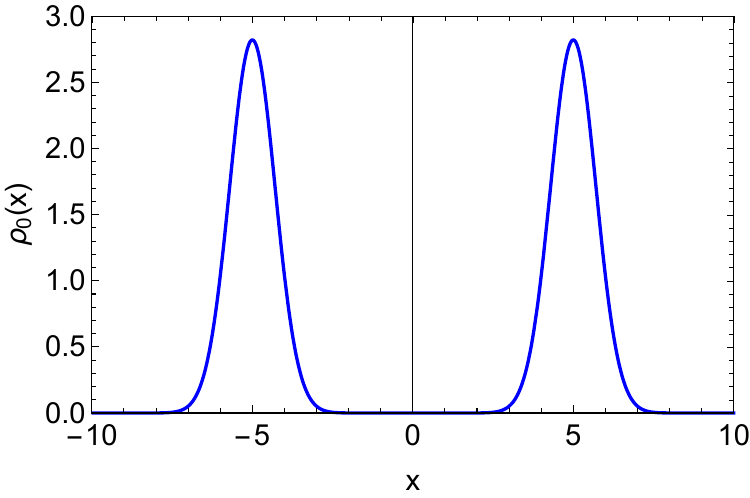} }
   \subfigure[\ $V(x)$]
   { \label{f:fig1b}
     \includegraphics[width=0.9\linewidth]{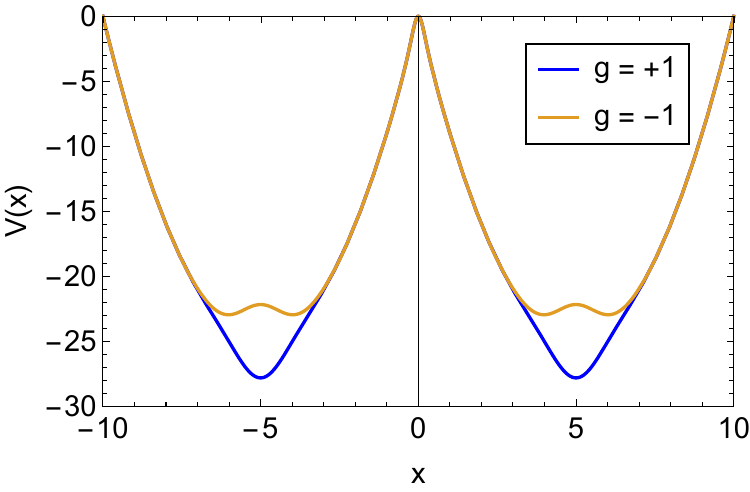} }
   \caption{\label{f:fig1} Plot of the density $\rho_0(x)$ (top) and
   confining potential $V(x)$ (bottom) both as functions of $x$ for
   $g = \pm 1$, and for the case when
   $a=1$, $q=5$, and $M_0 = 10$.  The chemical potential is
   $\mu_0 = a (1 - a q^2) = -24$.}
\end{figure}

Let us first choose for our two-trapped soliton wave function, the sum of two
Gaussians in 1D. For this case, the solution $u_0(x)$ is given by
\begin{align}
   u_0(x)
   &=
   A_0 \, \Bigl [\, \rme^{- a(x - q)^2/2} + \rme^{- a(x + q)^2/2 }\, \Bigr ]
   \label{e.2S1D:u0def} \\
   &=
   2 A_0 \, \rme^{-a(q^2 + x^2)/2} \, \cosh(a q x) \>.
   \notag
\end{align}
The conserved mass follows from Eq.~\eqref{eq:cons_mass}, and gives
\begin{equation}\label{e.2S1D:M0def}
   M_0
   =
   \int_{-\infty}^{\infty} \!\!\! \dd{x} u_0^2(x)
   =
   2 \sqrt{\frac{\pi}{a}} \bigl ( 1 + \rme^{-a q^2} \bigr ) \, A_0^2 \>,
\end{equation}
with the respective confining potential [cf. Eq.~\eqref{e:Vdef}]
given by
\begin{subequations}\label{e:V0VVdefs}
\begin{align}
   V(x) 
   &= 
   V_0(x) - g \, u_0^2(x) \>,
   \label{e:VVdef} \\
   V_0(x)
   &
   =  \mu_0 + u''_0(x)/u_0(x)
   \label{e:V0def1} \\
   &= 
   a^2 x \, (\, x - 2 q \tanh(a q x) \,) \>,
   \notag
\end{align}
\end{subequations}
where we have chosen $\mu_0 = a (1 - a q^2)$ so that $V_0(0) = 0$.
[The primes in Eq.~\eqref{e:V0def1} stand for differentiation
wrt $x$.] Plots of the density $\rho_0(x) = u_0^2(x)$ and the confining
potential $V(x)$ are shown in the top and bottom panels of Fig.~\ref{f:fig1}
as functions of $x$ with parameter values $a=1$, $q=5$, and $M_0 = 10$ for
$g = \pm 1$. We note that we have set the chemical potential $\mu_0 = -24$
so that $V(0) = 0$ therein. It can be discerned from the bottom panel of the
figure that for this two-soliton ansatz, $V(x)$ consists of two near harmonic
wells located at $x = \pm q$ when $g=1$. On the other hand, and for $g=-1$,
the potential contains two double-well potentials whose local maxima are
located similarly at $x = \pm q$.

%
%
%
%

\begin{figure}[t]
   \centering
   \includegraphics[width=0.9\linewidth]{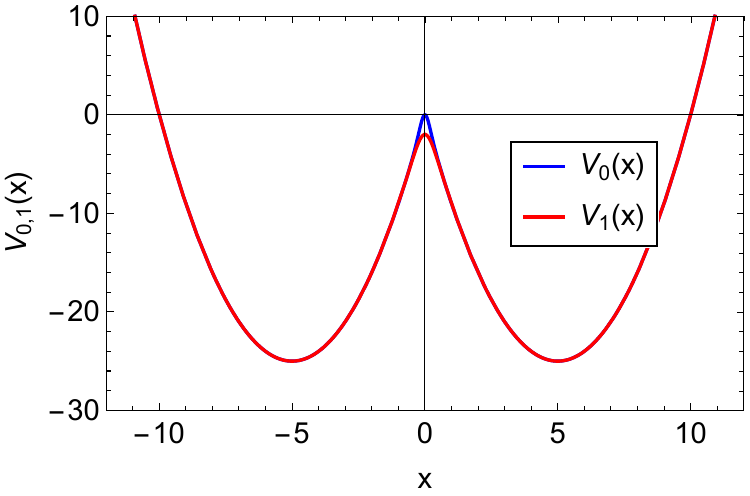}
   \caption{\label{f:fig2}Plots of $V_0(x)$ and $V_1(x)$ both as functions of $x$
   for the even and
   odd solitons that are given by Eqs.~\eqref{e:V0def1} and ~\eqref{e:VV0def},
   respectively.}
\end{figure}

The odd two-Gaussian soliton, defined by
\begin{align}\label{e.2S1D:u1def}
   u_1(x)
   &=
   A_1 \, \Bigl [\, \rme^{- a(x - q)^2/2} - \rme^{- a(x + q)^2/2 }\, \Bigr ]
   \notag \\
   &=
   2 A_1 \, \rme^{-a(q^2 + x^2)/2} \, \sinh(a q x)
\end{align}
with conserved mass
\begin{equation}\label{e.2S1D:M1def}
   M_1
   =
   \int_{-\infty}^{\infty} \!\!\! \dd{x} u_1^2(x)
   =
   2 \sqrt{\frac{\pi}{a}} \bigl ( 1 - \rme^{-a q^2} \bigr ) \, A_1^2 \>,
\end{equation}
and confining potential given by
\begin{subequations}\label{e:V1VV1defs}
\begin{align}
   V(x) 
   &= 
   V_1(x) - g \, u_1^2(x) \>,
   \label{e:VV1def} \\
   V_1(x)
   &
   =  \mu_0 + u''_1(x)/u_1(x)
   \label{e:VV0def} \\
   &= 
   a^2 x \, (\, x - 2 q \coth(a q x) \,) \>,
   \notag
\end{align}
\end{subequations}
has nearly the same soliton density distribution for these parameters,
and only a slightly different confining potential. Indeed, we compare
$V_0(x)$ (even soliton) and $V_1(x)$ (odd soliton) in Fig.~\ref{f:fig2}
[see, also, Eqs.~\eqref{e:V0def1} and ~\eqref{e:VV0def}] which showcases
that the only difference between them is the behavior near the origin.
An experimenter would be hard pressed to construct potentials which would
distinguish between even and odd solitons. The linear combinations 
\begin{equation}\label{e:upum}
   u_{\pm}(x)
   =
   [\, u_0(x) \pm u_1(x) \,]/2 \>,
\end{equation}
represent \emph{single} solitons at $x = \pm q$. The creation of two solitons
involves tunneling between the two harmonic wells. Similar results can be
obtained by using $\sech[ a(q \pm x) ]$ functions rather than Gaussian ones
to construct two soliton densities.
%
%
%
%

\begin{figure}[t]
   \centering
   \subfigure[\ $g = +1$]
   { \label{f:fig3a}
     \includegraphics[width=0.9\linewidth]{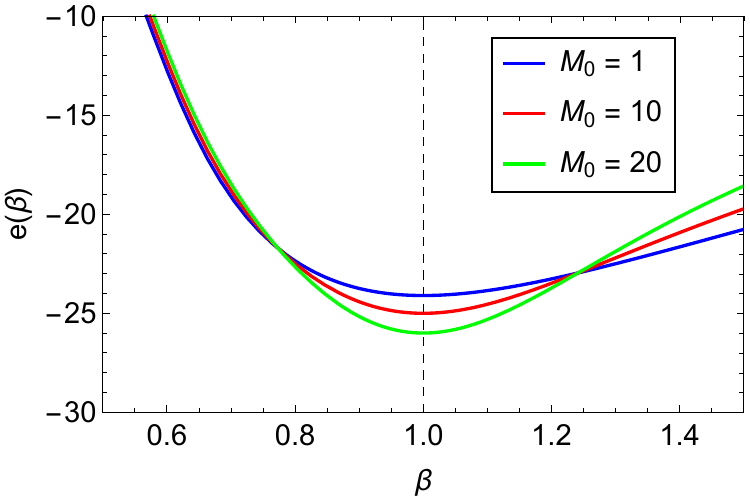} }
   \subfigure[\ $g = -1$]
   { \label{f:fig3b}
     \includegraphics[width=0.9\linewidth]{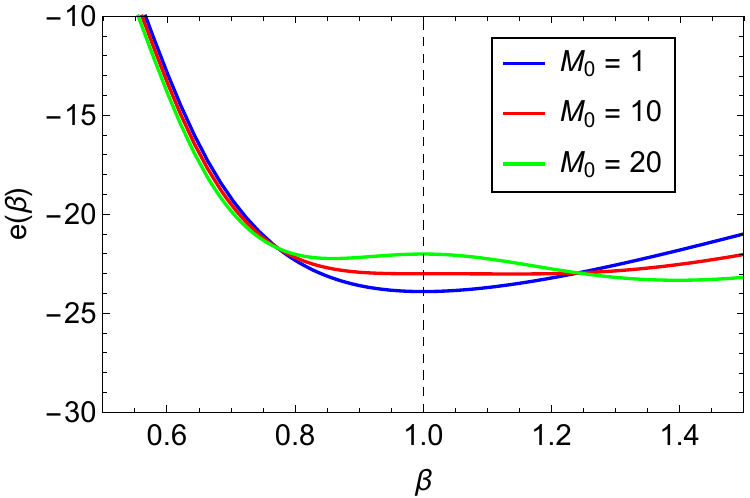} }
   \caption{\label{f:fig3} Plots of energy $e(\beta)$ vs.~$\beta$
   for $g = \pm 1$ for the case when $a=1$ and $q=5$. Note that the soliton
   for $g=1$ is always stable whereas for $g=-1$, it becomes unstable
   for $M_0 \approx 10$.}
\end{figure}

Stability of these solutions with regard to width stretching can be studied using
Derrick's theorem~\cite{Derrick-1964}. This theorem states that if the energy is
a minimum under the transformation $x \rightarrow \beta x$, i.e., dilation, keeping
the mass constant, the soliton is stable. The stretched wave function for the Gaussian
case $u_0(x)$ then becomes:
\begin{equation}\label{e:ustretched}
   u_{s}(x)
   =
   2 A_s \, \rme^{-a[q^2 + (\beta x)^2 ]/2} \, \cosh(a q \beta x) \>,
\end{equation}
where now the mass is given by
\begin{equation}\label{ee.2S1D:Ms}
   M_0
   =
   \int_{-\infty}^{\infty} \!\!\! \dd{x} u_s^2(x)
   =
   \frac{2}{\beta} \, \sqrt{\frac{\pi}{a}} \bigl ( 1 + \rme^{-a q^2} \bigr ) \, A_s^2 \>.
\end{equation}
Defining $e_i(\beta)\coloneqq E_i(\beta)/M_0$, the energy~\eqref{e:ConsE} is
then the sum of three terms: $e(\beta) = e_1(\beta) + e_2(\beta) + e_3(\beta)$, where
\begin{subequations}\label{e:Evalues}
\begin{align}
   e_1(\beta)
   &=
   \frac{1}{M_0} \int \dd{x} \, u^{\prime\, 2}_{s}(x)
   =
   \frac{a \beta^2}{2} \, 
   \Bigl [\,
      1 - \frac{2 a q^2}{1 + \rme^{a q^2}} \,
   \Bigr ]\>,
   \label{e:e1} \\
   e_2(\beta)
   &=
   \frac{g}{2 M_0} \int \dd{x} \, u^{4}_{s}(x)
   \label{e:e2} \\
   &=
   \frac{g M_0 \, \beta \sqrt{a}}{4 \sqrt{2 \pi}} \,
   \frac{( 4 \rme^{a q^2/2} + \rme^{2 a q^2} + 3 )}
        { ( 1 + \rme^{a q^2} )^2 } \>,
   \notag \\
   e_3(\beta)
   &=
   \frac{1}{M_0} \int \dd{x} \, V(x) \, u^{2}_{s}(x)
   \label{e:e3} \\
   &=
   \frac{1}{M_0} \int \dd{x} \, 
      [\, V_0(x) - g \, u_0^2(x) \,] \, u^{2}_{s}(x) \>,
   \notag
\end{align}
\end{subequations}
with $V_0(x)$ being given by \eqref{e:V0def1}. We note in passing that unlike the integrals
in Eqs.~\eqref{e:e1} and~\eqref{e:e2} which are evaluated explicitly, the integral in
Eq.~\eqref{e:e3} must be evaluated numerically. The top and bottom panels of Fig.~\ref{f:fig3}
depict the energy $e(\beta)$ as a function of $\beta$ for $g=1$ (top panel) and $g=-1$ (bottom panel),
respectively, with parameter values $a=1$ and $q=5$, and for several values of $M_0$. It can
be discerned from the top panel corresponding to the repulsive case (i.e., $g=1$) that at
$\beta=1$, the soliton is always stable for all values of $M_{0}$, however for the attractive
case (i.e., $g=-1$), the soliton becomes unstable for $M_0 \approx 10$.
%
%
%
%

\begin{figure}[t]
   \centering
   \subfigure[\ $g = +1$]
   { \label{f:fig4a}
     \includegraphics[width=0.9\linewidth]{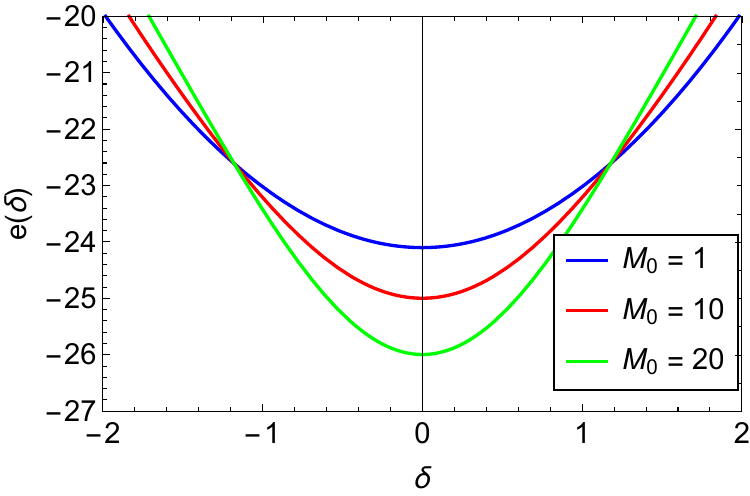} }
   \subfigure[\ $g = -1$]
   { \label{f:fig4b}
     \includegraphics[width=0.9\linewidth]{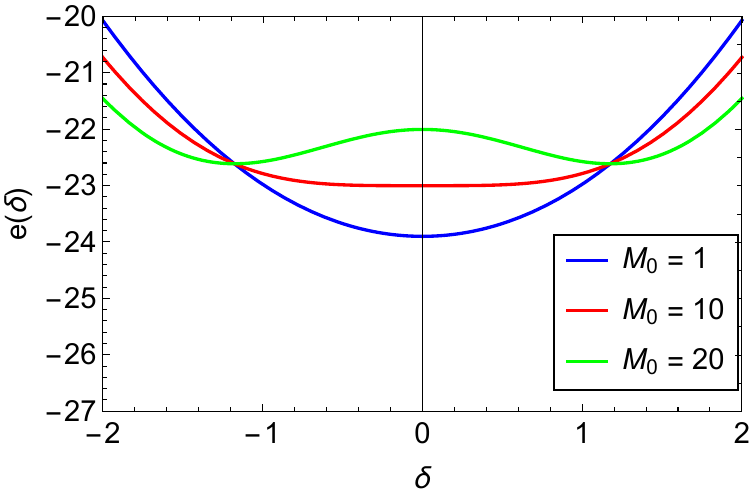} }
   \caption{\label{f:fig4} Same as Fig.~\ref{f:fig3} but for translational
   stability. Plots $e(\delta)$ vs.~$\delta$ for $g=1$ (top panel) and
   $g=-1$ (bottom panel) for the cases when $a=1$ and $q=5$. Note again
   that the soliton for $g=1$ is always stable whereas for $g=-1$, it
   becomes unstable for $M_0 \approx 10$.}
\end{figure}

Translational stability can be studied by displacing the soliton
solution $u_0(x)$ through the use of the transformation:
$x \rightarrow x \pm \delta$. In this case, the trial wave function
takes the form:
\begin{align}\label{e:utrans}
   u_t(x)
   &=
   A_t \, \Bigl [\, \rme^{- a[x - \delta - q]^2/2} + \rme^{- a(x + \delta + q)^2/2 }\, \Bigr ]
   \notag \\
   &=
   2 A_t \, \rme^{-a[(q + \delta)^2 + x^2]/2} \, \cosh[a ( q + \delta) x] \>,
\end{align}
where the mass is now given by
\begin{equation}\label{e.2S1D:Ms}
   M_0
   =
   \int_{-\infty}^{\infty} \!\!\! \dd{x} u_t^2(x)
   =
   2 \sqrt{\frac{\pi}{a}} \bigl ( 1 + \rme^{-a (q + \delta)^2} \bigr ) \, A_t^2 \>.
\end{equation}
Similarly, the energy is the sum of the following terms:
\begin{subequations}\label{e:Etvalues}
\begin{align}
   e_1(\delta)
   &=
   \frac{1}{M_0} \int \dd{x} \, u^{\prime\, 2}_{t}(x)
   =
   \frac{a}{2} \, 
   \Bigl [\,
      1 - \frac{2 a (q + \delta)^2}{1 + \rme^{a (q + \delta)^2}} \,
   \Bigr ]\>,
   \label{e:et1} \\
   e_2(\delta)
   &=
   \frac{g}{2 M_0} \int \dd{x} \, u^{4}_{t}(x)
   \label{e:et2} \\
   &=
   \frac{g M_0 \sqrt{a}}{4 \sqrt{2 \pi}} \,
   \frac{( 4 \rme^{a (q + \delta)^2/2} + \rme^{2 a (q + \delta)^2} + 3 )}
        { ( 1 + \rme^{a (q + \delta)^2} )^2 } \>,
   \notag \\
   e_3(\delta)
   &=
   \frac{1}{M_0} \int \dd{x} \, V(x) \, u^{2}_{t}(x)
   \label{e:et3} \\
   &=
   \frac{1}{M_0} \int \dd{x} \, 
      [\, V_0(x) - g \, u_0^2(x) \,] \, u^{2}_{t}(x) \>,
   \notag
\end{align}
\end{subequations}
where $V_0(x)$ is given by \eqref{e:V0def1}. Again, this last integral must
be evaluated numerically. We note in passing that the energy components of
Eqs.~\eqref{e:et1}-\eqref{e:et3} can be respectively obtained from
Eqs.~\eqref{e:e1}-\eqref{e:e3} upon setting $\beta=1$ and replacing $q\mapsto q+\delta$.
We plot the energy $e(\delta)$ as a function of $\delta$ in Fig.~\ref{f:fig4}
for the case with $a=1$ and $q=5$, and for several values of $M_0$ (again, for
both $g=1$ and $g=-1$). At $\delta=0$, and for the repulsive case ($g=1$), the
soliton is always stable for all values of $M_0$, however for the attractive
case ($g=-1$), the soliton again becomes unstable for $M_0 \approx 10$.
%
%
%
%

Based on the above two variational studies, we conclude that
most likely the two soliton solutions are always stable for repulsive
case ($g=1$) but become unstable for the attractive case ($g=-1$).
We have also studied a two-soliton wave functions of the form:
 $u_0(x,y) = A \, [\, \sech(q - x) + \sech(q + x) \,]$, 
which gives a similar density distribution as the Gaussian case.
Numerical results for stretching and translational stability for
this ansatz are similar to the Gaussian case discussed above,
and indicate stability for the repulsive case and instability
for $M \gtrsim 10$ for the attractive case.  We will not present
those results here.

%
%
\subsection{\label{ss:2D}Two dimensions}

%
%
\subsubsection{\label{sss:2D-case1}Case 1}
%
%
\begin{figure}[t]
   \centering
   \subfigure[\ $\rho_0(x,y)$]
   { \label{f:fig5a}
     \includegraphics[width=0.9\linewidth]{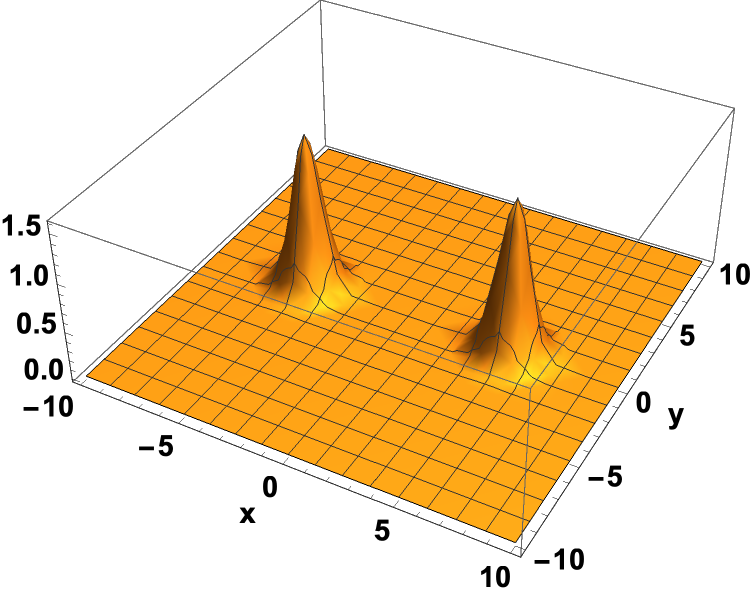} }
   \subfigure[\ $V_0(x,y)$]
   { \label{f:fig5b}
     \includegraphics[width=0.9\linewidth]{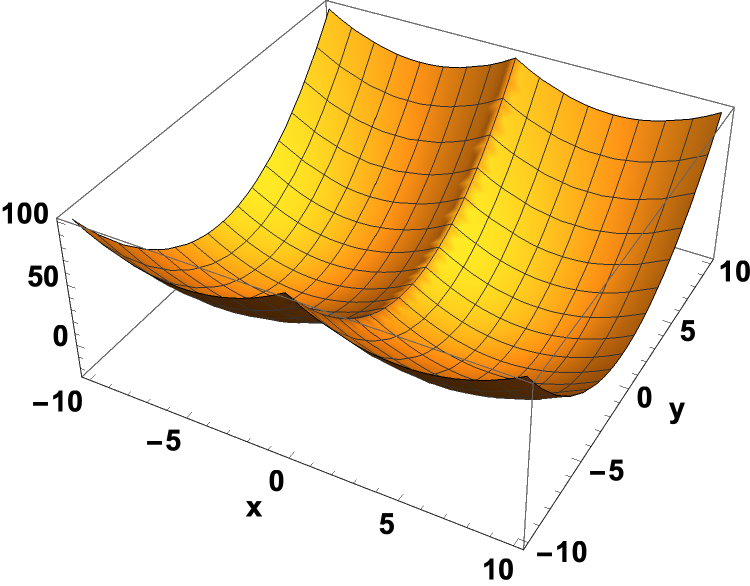} }
   \caption{\label{f:fig5} (a) Plots of the density $\rho_0(x,y)$ and
   (b) confining potential $V_0(x,y)$ as functions of $x$ and $y$ for
   the case when $a=b=1$, $q=5$, and $M_0 = 10$.  The chemical potential
   is $\mu_0 = b + 2 \sech^2(a) - 1 = 0.000363$.}
\end{figure}
%
%
We proceed next with the construction of a 2D wave function
consisting of two Gaussian functions. In particular, we assume a
Gaussian in the $x$ direction centered at $x = \pm q$, and one in
the $y$ direction centered at $y=0$. The ansatz we consider
is given explicitly by:
\begin{align}
   u_0(x,y)
   &=
   A_0 \, 
   \bigl \{\,
      \rme^{- [a(x - q)^2 + b y^2]/2} + \rme^{- [a(x + q)^2 + b y^2]/2} \, 
   \bigr \}  
   \label{e.2D:u0} \\
   &=
   2 A_0 \, \rme^{- [a(x^2 + q^2) + b y^2]/2} \, \cosh(a q x) \>.
   \notag
\end{align}
For this case, the conserved mass is given by
\begin{equation}\label{e.2D:M0}
   M_0
   =
   \frac{2\pi}{\sqrt{a b}} ( 1 + \rme^{-a q^2} ) \, A_0^2 \>,
\end{equation}
and the confining potential by:
\begin{subequations}\label{e.2D:V0VVdefs}
\begin{align}
   V(x,y) 
   &= 
   V_0(x,y) - g \, u_0^2(x,y) \>,
   \label{e:VVdef2} \\
   V_0(x,y)
   &
   =  \mu_0 + \{\, [ \partial_x^2  + \partial_y^2 ] u_0(x,y) \,\}/u_0(x,y)
   \label{e:V0deff2} \\
   &= 
   a^2 x^2 + b^2 y^2 - 2 a^2 q x \tanh(a q x) \>,
   \notag
\end{align}
\end{subequations}
where we have chosen $\mu_0 = a+b-(a q)^2$ so that $V_0(0,0) = 0$.
Plots of the density $\rho_0(x,y) = u_0^2(x,y)$ and the potential
$V_0(x,y)$ (both as functions of $x$ and $y$) for the case when
$a=b=1$, $q=5$, and $M_0 = 10$ are shown in Fig.~\ref{f:fig5}.

\begin{figure}[t]
   \centering
   \subfigure[\ $g = +1$]
   { \label{f:fig6a}
     \includegraphics[width=0.9\linewidth]{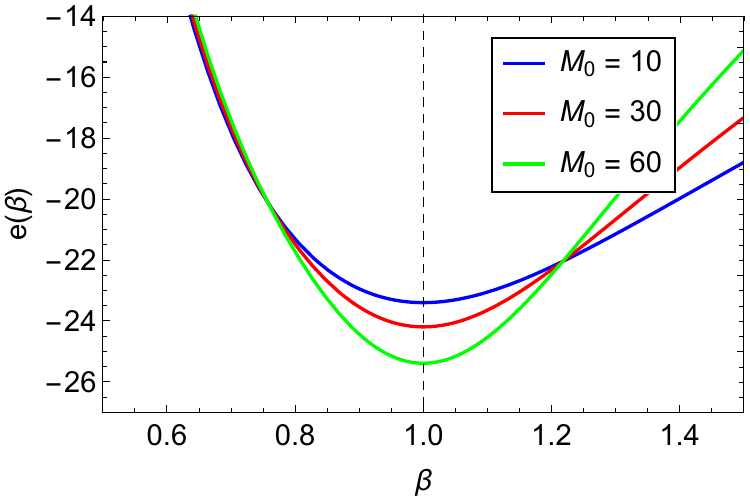} }
   \subfigure[\ $g = -1$]
   { \label{f:fig6b}
     \includegraphics[width=0.9\linewidth]{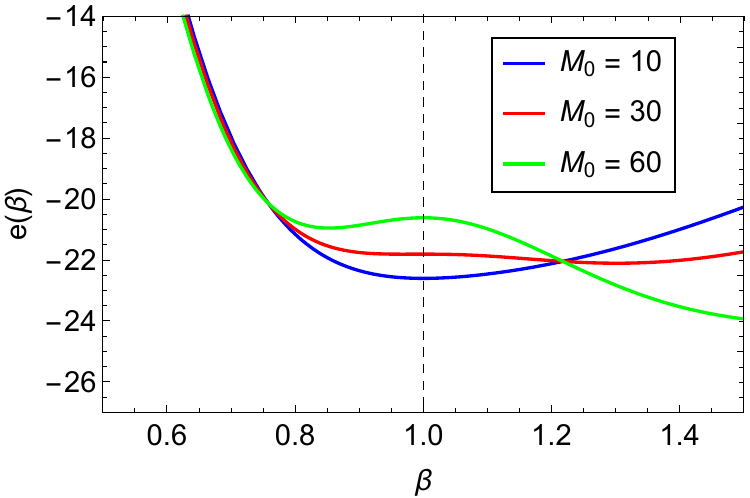} }
   \caption{\label{f:fig6} Plots of the total energy $e(\beta)$
   vs. $\beta$ for $g = 1$ (top panel) and $g=-1$ (bottom
   panel), and for case 1 when $a=b=1$ and $q=5$.}
\end{figure}

To study stability with respect to a stretching of the coordinates
$x \rightarrow \beta x$ and $y \rightarrow \beta y$, we use a trial
wave function of the form:
\begin{equation}\label{e.2D:us}
   u_s(x,y)
   =
   2 A_s \, \rme^{-  [a((\beta x)^2 + q^2) + b (\beta y)^2]/2} \, \cosh(a q \beta x) \>,
\end{equation}
where now the mass is given by
\begin{equation}\label{e.2D:Ms}
   M_0
   =
   \int\!\! \dd[2]{x} u_s^2(x,y)
   =
   \frac{2 \pi}{\beta^2 \sqrt{ab}} \bigl ( 1 + \rme^{-a q^2} \bigr ) \, A_s^2 \>.
\end{equation}
Again computing components of the energy under stretching,
we find:
\begin{subequations}\label{e.2D:Derrick-e}
\begin{align}
   e_1(\beta)
   &=
   \frac{\beta^2}{2} \, 
   \Bigl [\,
      a + b - \frac{2 a^2 q^2}{1 + \rme^{a q^2}} \,
   \Bigr ] \>,
   \label{e.D2:Derrick-e1} \\
   e_2(\beta)
   &=
   \frac{g M_0 \, \beta^2 \sqrt{a b}}{16 \pi} \,
   \frac{( 8 \rme^{a q^2/2} + 2 \rme^{2 a q^2} + 6 )}
        { ( 1 + \rme^{a q^2} )^2 } \>,
   \label{e.D2:Derrick-e2} \\
   e_3(\beta)
   &=
   \frac{1}{M_0} \int\!\! \dd[2]{x} \, V(x,y) \, u^{2}_{s}(x,y)
   \label{e.D2:Derrick-e3} \\
   &=
   \frac{1}{M_0} \int\!\! \dd[2]{x} 
      [\, V_0(x,y) - g \, u_0^2(x,y) \,] \, u^{2}_{s}(x,y) \>,
   \notag   
\end{align}
\end{subequations}
where the integral in Eq.~\eqref{e.D2:Derrick-e3} has to be evaluated
numerically. The total energy $e(\beta) = e_1(\beta)+e_2(\beta)+e_3(\beta)$
is presented in Fig.~\ref{f:fig6} as a function of $\beta$ for $g=\pm 1$ for
the case when $a=b=1$ and $q=5$, and for various values of the mass $M_0$.
%
%
%
%
%
%
\begin{figure}[t]
   \centering
   \subfigure[\ $g = +1$]
   { \label{f:fig7a}
     \includegraphics[width=0.9\linewidth]{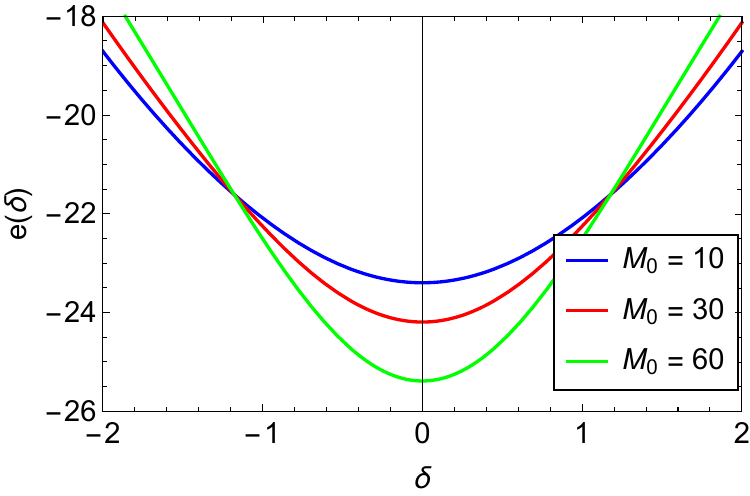} }
   \subfigure[\ $g = -1$]
   { \label{f:fig7b}
     \includegraphics[width=0.9\linewidth]{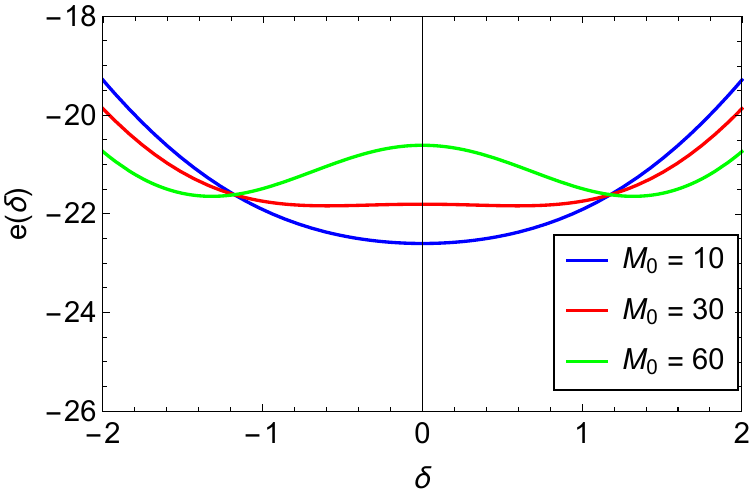} }
   \caption{\label{f:fig7} Same as Fig.~\ref{f:fig6} but for translational
   stability. Plots $e(\delta)$ vs.~$\delta$ for $g=1$ (top panel) and
   $g=-1$ (bottom panel) for case 1 when $a=b=1$ and $q=5$.
}
\end{figure}
%
%
It can be discerned from the figure that at $\beta=1$, the soliton
for the repulsive case (i.e., $g=1$) is always stable for all values
of $M_0$, whereas for the attractive  case ($g=-1$), the soliton remains
stable for values of $M_0 \lesssim 30$ but becomes unstable for larger
values of $M_0$.

Translational stability is studied by making the replacement $q \rightarrow q + \delta$,
and computing the energy as a function of $\delta$. The trial wave function
in this case is given by
\begin{equation}\label{e.2D:Trans-WF}
   u_t(x,y)
   =
   2 A_t \, \rme^{- [a(x^2 + (q+\delta)^2) + b y^2]/2} \, \cosh[a (q + \delta) x] \>,
\end{equation}
where the total mass reads
\begin{equation}\label{e.2D:At}
   M_0
   =
   \int\!\! \dd[2]{x} u_t^2(x,y)
   =
   \frac{2 \pi}{\sqrt{ab}} \bigl ( 1 + \rme^{-a (q+\delta)^2} \bigr ) \, A_t^2 \>.
\end{equation}
Same as before, the energy terms in this case, i.e., for translational
stability are obtained from Eqs.~\eqref{e.2D:Derrick-e} by setting $\beta=1$
followed by the replacement $q\mapsto q+\delta$.
The results for this case are shown in
Fig.~\ref{f:fig7} where the energy $e(\delta)$ is plotted against
$\delta$ for $g=1$ and $g=-1$ (see the top and bottom panels, respectively).
The soliton solutions for $g=1$ solitons are always stable whereas the
ones with $g=-1$ are stable for values of mass $M_0 \lesssim 30$,
and become unstable for larger values of the mass, in agreement
with the results of Derrick's theorem in Fig.~\ref{f:fig6}.

%
%
\subsubsection{\label{sss:2D-case2}Case 2}
%
%
\begin{figure}[t]
   \centering
   \subfigure[\ $\rho_0(x,y)$]
   { \label{f:fig52a}
     \includegraphics[width=0.9\linewidth]{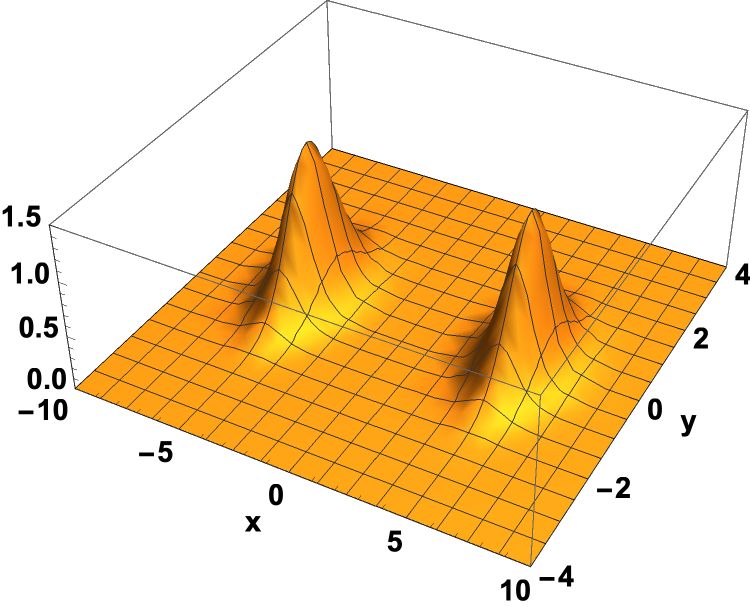} }
   \subfigure[\ $V_0(x,y)$]
   { \label{f:fig52b}
     \includegraphics[width=0.9\linewidth]{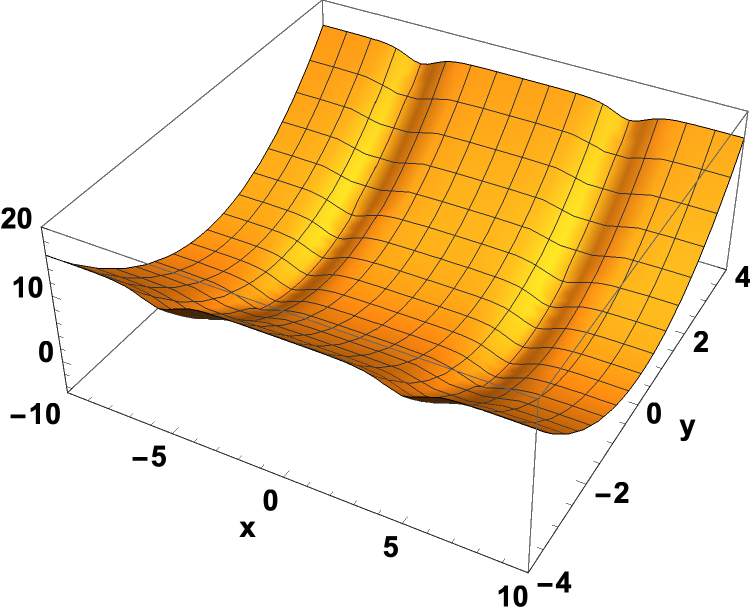} }
   \caption{\label{f:fig52} (a) Plots of the density $\rho_0(x,y)$ and
   (b) the confining potential $V_0(x,y)$ (again, as functions of $x$ and $y$)
   for the case when $b=1$, $q=5$,
   and $M_0 = 10$.  The chemical potential is
   $\mu_0 = b + 2 \sech^2(q) - 1 = 0.000363$.}
\end{figure}

In this case we construct a 2D wave function consisting of two $\sech(x \pm q)$
functions centered at $x = \pm q$, and a Gaussian in the $y$ direction centered
at $y=0$.  Explicitly we choose:
\begin{equation}\label{e.2D:u02}
   u_0(x,y)
   =
   A_0 \, [\, \sech(x+q) + \sech(x-q) \,] \, \rme^{ - b y^2/2} \>.
\end{equation}
For this case, the conserved mass is given by
\begin{equation}\label{e.2D:M02}
   M_0
   =
   4 \sqrt{\frac{\pi}{b}} [\,1 + q \csch(q) \sech(q) \,] \, A_0^2 \>,
\end{equation}
and the confining potential by:
\begin{subequations}\label{e.2D:V0VVdefs2}
\begin{align}
   V(x,y) 
   &= 
   V_0(x,y) - g \, u_0^2(x,y) \>,
   \label{e:VVdef23} \\
   V_0(x,y)
   &=  
   \mu_0 + \{\, [ \partial_x^2  + \partial_y^2 ] u_0(x,y) \,\}/u_0(x,y)
   \label{e:V0def} \\
   &= 
   b^2 y^2 + 2 \sech^2(q) - 2 [\,\sech^2(q-x) 
   \notag \\
   & \hspace{1em}
   - \sech(q-x) \sech(q+x) + \sech^2(q+x) \,] \>,
   \notag
\end{align}
\end{subequations}
where we have chosen $\mu_0 = b + 2 \sech^2(q) - 1$ so that $V_0(0,0) = 0$.
Plots of the density $\rho_0(x,y) = u_0^2(x,y)$ and the potential
$V_0(x,y)$ as functions of $x$ and $y$ for the case when $b=1$, $q=5$,
and $M_0 = 10$ are shown in Fig.~\ref{f:fig52}.

\begin{figure}[t]
   \centering
   \subfigure[\ $g = +1$]
   { \label{f:fig62a}
     \includegraphics[width=0.9\linewidth]{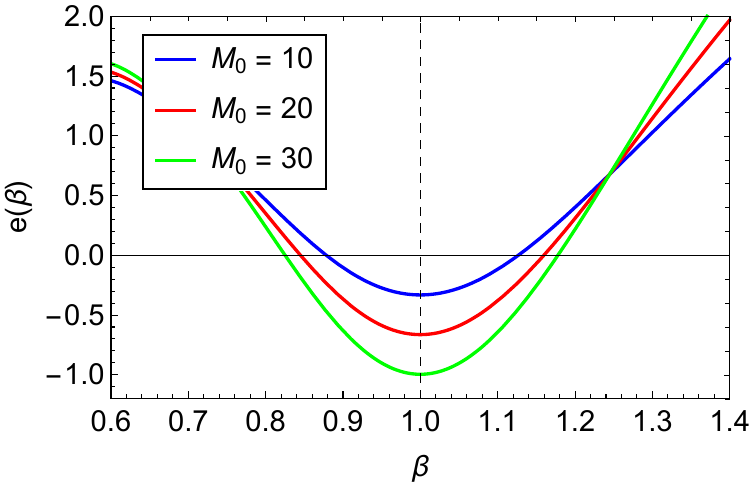} }
   \subfigure[\ $g = -1$]
   { \label{f:fig62b}
     \includegraphics[width=0.9\linewidth]{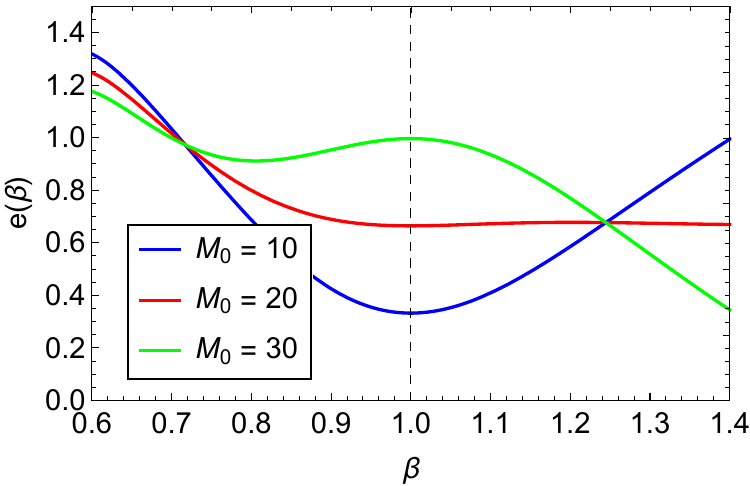} }
   \caption{\label{f:fig62} Plots of the total energy $e(\beta)$
   vs. $\beta$ for $g = 1$ (top panel) and $g=-1$ (bottom
   panel), and for case 2 when $b=1$ and $q=5$.}
\end{figure}

The stability with respect to a stretching of the coordinates
$x \rightarrow \beta x$ and $y \rightarrow \beta y$ is
carried out by assuming the trial wave function:
\begin{equation}\label{e.2D:us2}
   u_s(x,y)
   =
   A_s \, [\, \sech(\beta x+q) + \sech(\beta x-q) \,] \, \rme^{ - b \beta^2 y^2/2} \>,
\end{equation}
where now the mass is given by
\begin{equation}\label{e.2D:Ms2}
   M_0
   =
   \frac{4}{\beta^2} \sqrt{\frac{\pi}{b}} [\,1 + q \csch(q) \sech(q) \,] \, A_0^2 \>.
\end{equation}
Upon computing the energy components in this case we find
\begin{subequations}\label{e.2D:Derrick-e2}
\begin{align}
   e_1(\beta)
   &=
   \frac{\beta^2}{6[1 + q \csch(q) \sech(q)]} \, 
   \label{e.D2:Derrick-e1-2} \\
   & \hspace{-1em}
   \times
   \bigl \{\,
      2 + 3b + 12 \coth(2q) \csch(2q) 
      \notag \\
      &\hspace{1em}
      - 3q[ 6 + b +(2 - b)\cosh(4q) \csch^3(2q)]
   \bigr \} \>,
   \notag \\
   e_2(\beta)
   &=
   \frac{g M \beta^2}{96} \sqrt{\frac{b}{2 \pi}} \,
   \frac{\csch(q) \sech(q)}{[2 q + \sinh(2q)]^2}
   \label{e.D2:Derrick-e2-2} \\
   & \hspace{-1em}
   \times
   \bigl \{
      -48 q + 72 q \cosh(2q) - 39 \sinh(2q) 
      \notag \\
      & \hspace{1em}
      + 12 \sinh(4q) + \sinh(6q) \,
   \bigr \} \>,
   \notag \\
   e_3(\beta)
   &=
   \frac{1}{M_0} \int\!\! \dd[2]{x} \, V(x,y) \, u^{2}_{s}(x,y)
   \label{e.D2:Derrick-e3-2} \\
   &=
   \frac{1}{M_0} \int\!\! \dd[2]{x} 
      [\, V_0(x,y) - g \, u_0^2(x,y) \,] \, u^{2}_{s}(x,y) \>,
   \notag   
\end{align}
\end{subequations}
where the integral in Eq.~\eqref{e.D2:Derrick-e3-2} has to be evaluated
numerically. The total energy $e(\beta) = e_1(\beta)+e_2(\beta)+e_3(\beta)$
is presented in Fig.~\ref{f:fig62} as a function of $\beta$ for $g=\pm 1$ for
the case when $a=b=1$ and $q=5$, and for various values of the mass $M_0$.
%
%
%
%
%
%
\begin{figure}[t]
   \centering
   \subfigure[\ $g = +1$]
   { \label{f:fig72a}
     \includegraphics[width=0.9\linewidth]{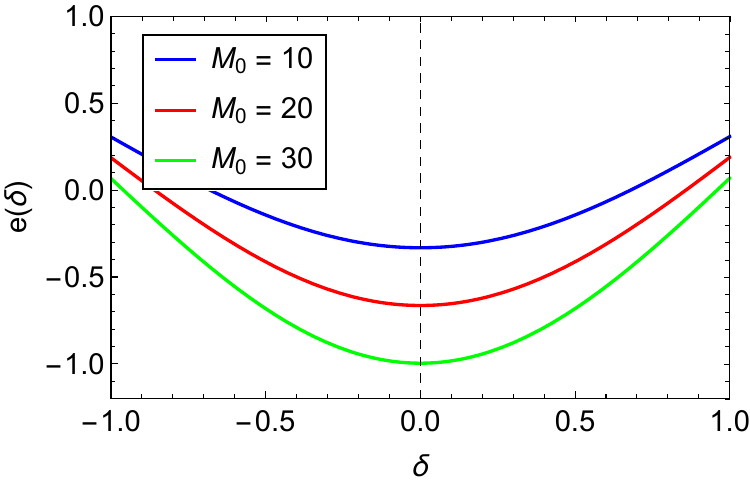} }
   \subfigure[\ $g = -1$]
   { \label{f:fig72b}
     \includegraphics[width=0.9\linewidth]{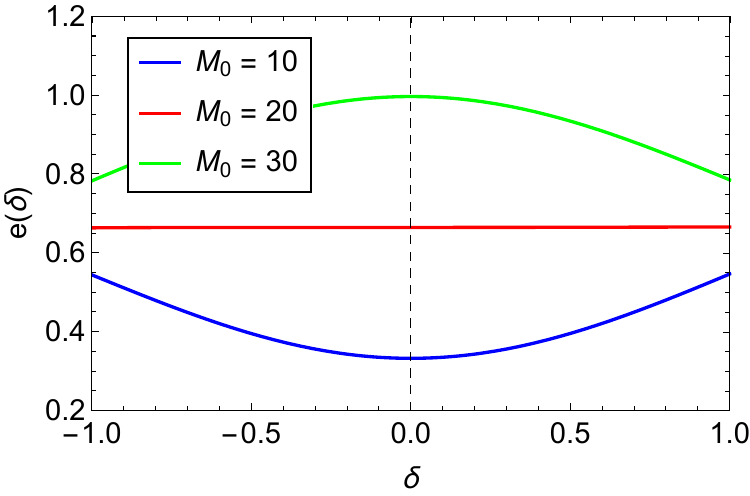} }
   \caption{\label{f:fig72} Same as Fig.~\ref{f:fig62} but for translational
   stability. Plots of $e(\delta)$ vs.~$\delta$ for $g=1$ (top panel) and
   $g=-1$ (bottom panel) for the case 2 when $b=1$ and $q=5$.}
\end{figure}
%
%
It can be discerned from the figure that at $\beta=1$, the soliton
for the repulsive case (i.e., $g=1$) is always stable for all values
of $M_0$, whereas for the attractive  case ($g=-1$), the soliton remains
stable for values of $M_0 \lesssim 20$ but becomes unstable for larger
values of $M_0$.

Translational stability is studied by making the replacement $q \rightarrow q + \delta$,
and computing the energy as a function of $\delta$. The trial wave function
in this case is given by
\begin{equation}\label{e.2D:Trans-WF2}
   u_t(x,y)
   =
   A_t \, [\, \sech(x+q+\delta) + \sech(x-q-\delta) \,] \, 
   \rme^{ - b y^2/2} \>,
\end{equation}
where the total mass is now given by
\begin{equation}\label{e.2D:At2}
   M_0
   =
   4 \sqrt{\frac{\pi}{b}} 
   [\,1 + (q+\delta) \csch(q+\delta) \sech(q+\delta) \,] \, A_t^2 \>.
\end{equation}
Again, the energy terms for translational instability are obtained from
the expressions~\eqref{e.2D:Derrick-e2} by setting initially $\beta=1$,
and making the replacement $q \rightarrow q + \delta$ afterwards.
The results in this case for the energy $e(\delta)$ as a function of $\delta$ are shown in
Fig.~\ref{f:fig72}. The $g=1$ solitons are always stable whereas the $g=-1$ solitons are stable
for values of mass $M_0 \lesssim 20$, and become unstable for larger values of the mass, in
agreement with the results of Derrick's theorem in Fig.~\ref{f:fig62}.

%
%
\subsection{\label{ss:3D}Three dimensions}

Two spheroidal BEC solitons have been studied for a variety of reasons in the
literature, the most intriguing being to determine whether modifications of
quantum mechanics due to general relativity can be seen in this type of system.
In most of these problems an approximate confining potential is used so that
questions of stability of the BEC as one increases the number of atoms can
be addressed. Indeed, we can first reverse engineer the exact potential needed to
make the sum of two Gaussians an exact solution. Then, we can determine the stability
criteria for soliton solutions using Derrick's theorem as well as linear response
theory.

%
%
\subsubsection{\label{sss:3D2S}Two solitons}

\begin{figure}[t]
   \centering
   \subfigure[\ $\rho_0(x,y,z)$]
   { \label{f:fig8a}
     \includegraphics[width=0.9\linewidth]{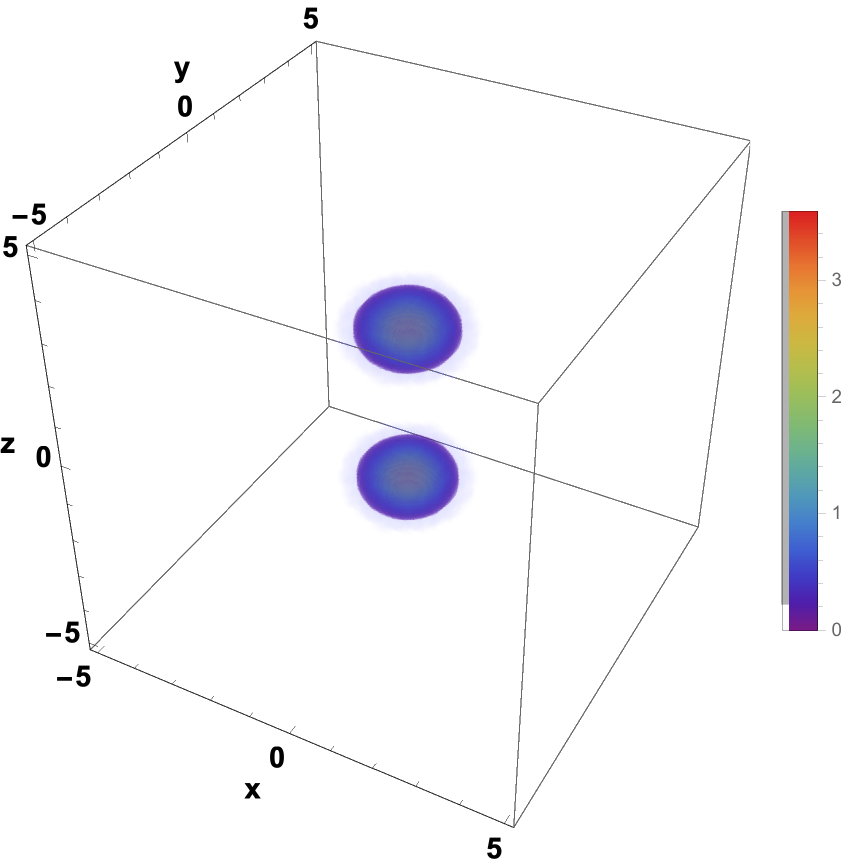} }
   \subfigure[\ $V_0(x,y,z)$]
   { \label{f:fig8b}
     \includegraphics[width=0.9\linewidth]{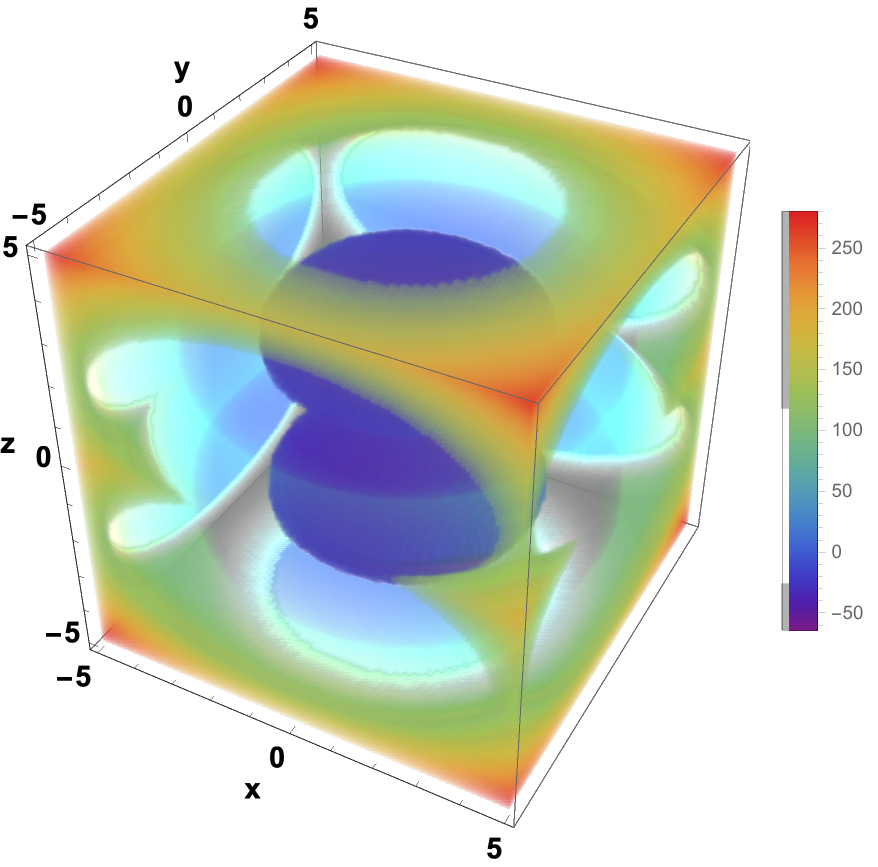} }
   \caption{\label{f:fig8} (a) Plots of the density $\rho_0(x,y,z)$ and
   (b) the confining potential $V_0(x,y,z)$ (both as functions of $x$, $y$, and
   $z$) for the case when $a=2$, $b=4$, $q=2$,
   and $M_0 = 10$.  The chemical potential is
   $\mu_0 = b + 2 a - (b q)^2 = -56$.}
\end{figure}

We start by constructing a 3D Gaussian, two-soliton solution
of the form:
\begin{align}
   u_0(x,y,z)
   &=
   A_0 \, \rme^{-a (x^2 + y^2)/2}
   \bigl [\,
      \rme^{ - b (q + z)^2/2 } + \rme^{ - b (q - z)^2/2 }\,
   \bigr ]
   \label{e.3D:u0} \\
   &=
   2 A_0 \, \rme^{- [a (x^2 + y^2) + b(z^2 + q^2)]/2} \cosh(b q z) \>.
   \notag
\end{align}
Here we chose the center of the soliton at $x = y = 0$ and $z= \pm q$
for simplicity. The mass is now given by:
\begin{equation}\label{e.3D:M0}
   M_0
   =
   \frac{2\pi^{3/2}}{a\sqrt{b}} ( 1 + \rme^{-b q^2} ) \, A_0^2 \>,
\end{equation}
and the confining potential by:
\begin{align}
   V(x,y,z)
   &=
   V_0(x,y,z) - g \, u_0^2(x,y,z) \>,
   \label{e.3D:V0} \\
   V_0(x,y,z)
   &=
   a^2 (x^2 + y^2) + b^2 z^2 - 2 b^2 q z \tanh(b q z) \>,
   \notag
\end{align}
where we have chosen $\mu_0 = b + 2a - b^2 q^2$ so that $V_0(0,0,0)=0$.
Plots of the density $\rho_0(x,y,z)$ and potential $V_0(x,y,z)$ as functions
of $x$, $y$, and $z$, are shown in Fig.~\ref{f:fig8}.
%
%
%
%

%
%
\subsubsection{\label{sss:3DsS}Three solitons}

\begin{figure}[t]
   \centering
   \subfigure[\ $\rho_0(x,y,z)$]
   { \label{f:fig9a}
     \includegraphics[width=0.8\linewidth]{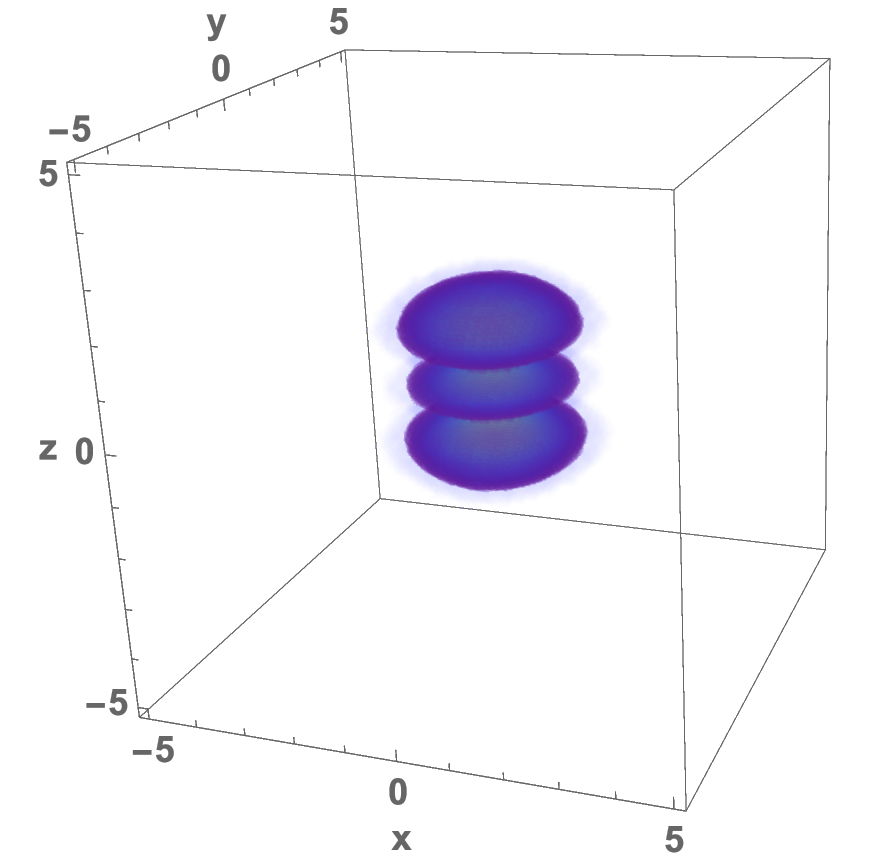} }
   \subfigure[\ $V_0(x,y,z)$]
   { \label{f:fig9b}
     \includegraphics[width=0.8\linewidth]{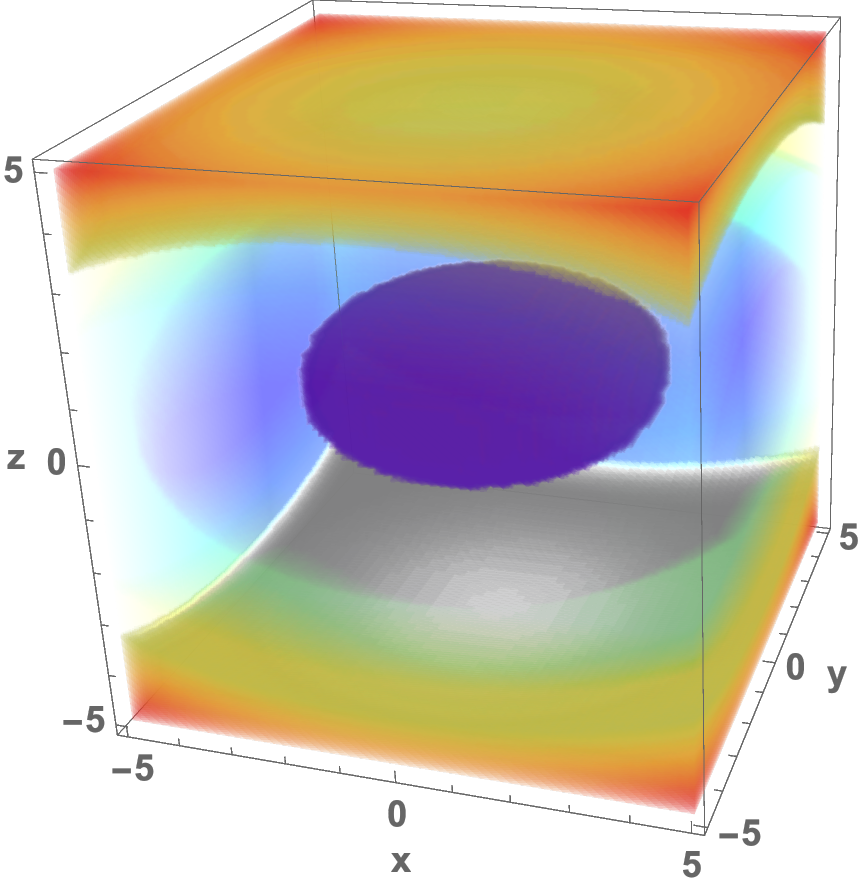} }
   \caption{\label{f:fig9} (a) Plots of the density $\rho_0(x,y,z)$ and
   (b) the confining potential $V_0(x,y,z)$ (both as functions of $x$, $y$,
   and $z$) for the three-soliton Hermite
   case when $n=2$, $a=1$, $b=2$, and $M_0 = 10$. The chemical potential is
   $\mu_0 = 2 a + 5 b = 12$.}
\end{figure}

There are many possibilities for obtaining $N$-soliton solutions
in 3D.  The simplest three soliton case is given by
%
\begin{equation}\label{e.H:u0}
   u_0(x,y,z)
   =
   A_0 \, \rme^{ -[a (x^2 + y^2)/2 + b z^2]/2 } \, H_n(\sqrt{b}\, z) \>,
\end{equation}
where $H_n(\zeta)$ is a Hermite polynomial of order $n$.
In this case, the conserved mass is given by
\begin{equation}\label{e.H:M0}
   M_0
   =
   \frac{\pi^{3/2}  \, 2^2 n!}{a \sqrt{b}} \, A_0^2 \>,
\end{equation}
and the confining potential by
\begin{align}
   V(x,y,z)
   &=
   V_0(x,y,z) - g \, u_0^2(x,y,z) \>,
   \label{e.H:Vdef} \\
   V_0(x,y,z)
   &=
   2 b n + a^2 (x^2 + y^2) + b^2 z^2
   \notag \\
   & \hspace{-2em}
   -
   \frac{4 b^{3/2} \, n z \, H_{n-1}(\sqrt{b}\,z) 
             - 4 b \, n (n-1) \, H_{n-2}(\sqrt{b}\,z)}
        { H_n(\sqrt{b}\, z) } \>,
   \notag
\end{align}
where we have chosen $\mu_0 = 2 a + (2 n + 1)\, b$ so that $V_0(0,0,0)=0$.
Plots of the density $\rho_0(x,y,z)$ and potential $V_0(x,y,z)$ (both as
functions of $x$, $y$, and $z$) are shown in Fig.~\ref{f:fig9} for the three soliton
case with parameter values $n=2$, $a=1$, $b=2$, and $M_0 = 10$.
%
%
%
%

One can analytically determine the energy of the stretched soliton in this case with
$x_i \rightarrow \beta x_i$ ($i=1,2,3$) keeping the mass $M$ fixed. The total
energy in this case is:
 \begin{equation} 
   e(\beta) =  e_1(\beta) + e_2(\beta) + e_3(\beta) \>,
\end{equation}
with
\begin{subequations}\label{e:Derrick3D}
\begin{align}
   e_1(\beta) 
   &=
   \frac{1}{2} \beta ^2 (2 a+5 b) \>,
   \label{e:Derrick3D-a} \\
   e_2(\beta) 
   &= 
   \frac{41 a \sqrt{b} \beta ^3 g M}{256 \sqrt{2} \pi ^{3/2}} \>,
   \label{e:Derrick3D-b} \\
   e_3(\beta) 
   &= 
   \frac{2 a+5 b}{2 \beta ^2}
   \label{e:Derrick3D-c} \\
   & \hspace{1em}
   -
   \frac{a \sqrt{b} \left(2 \beta ^8
   -
   16 \beta ^6+69 \beta ^4-16
   \beta ^2- 2\right) \beta ^3 g M}{4 \pi ^{3/2} \left(\beta ^2+1\right)^{11/2}} \>.
   \notag
\end{align}
\end{subequations}
One easily verifies that $\beta=1$ is a stationary point. Setting the
second derivative to zero at $\beta=1$ gives the criterion for instability
of the $g<0$ soliton to set in, that is
\begin{equation} 
   M_c 
   = 
   - g \frac{2048 \sqrt{2} \pi ^{3/2} (2 a+5 b)}{1527 a \sqrt{b}}\>.
\end{equation}
For $g=-1$, $a=2,b=2$ we find $M_c = 89.62$.

%
%
\section{\label{s:LinearResponse}Linear response equations}

Solutions of the linear response equations \eqref{e:DiffEq} are obtained by
consideration of an eigenvalue equation. Let the pair $(\, a(\vb{r}),b(\vb{r}) \,) \in \bbC^2$
satisfy the skew-symmetric eigenvalue equation:
\begin{equation}\label{e:eigenequ}
   \begin{pmatrix}
     [\, h(\vb{r})  + g u_0^2(\vb{r}) \,] & g u_0^2(\vb{r})  \\
     - g u_0^2(\vb{r})  & - [\, h(\vb{r})  + g u_0^2(\vb{r}) \,]
   \end{pmatrix}
   \begin{pmatrix}
      a(\vb{r}) \\
      b(\vb{r})
   \end{pmatrix}
   =
   \mu
   \begin{pmatrix}
      a(\vb{r}) \\
      b(\vb{r})
   \end{pmatrix} \>,
\end{equation}
where $\mu \in \bbC$ is the eigenvalue.  Here $h(\vb{r})$ is given by \eqref{e:hdef} 
\begin{equation}\label{e:hdef-II}
   h(\vb{r}) = -\laplacian + V_0(\vb{r}) \>,
\end{equation} 
and is \emph{independent} of the mass $M_0$.
Equations \eqref{e:eigenequ} are sometimes called the Bogoliubov-de Gennes (BdG) equations \cite{Bogolyubov-1947,deGennes-1966}.

By taking the complex conjugate of \eqref{e:eigenequ}, interchanging top and bottom
lines, and multiplying by $-1$, we see that if $(\, a(\vb{r}),b(\vb{r}) \,)$ are
a pair of solutions with eigenvalue $\mu$, then $(\, b^{\ast}(\vb{r}),a^{\ast}(\vb{r}) \,)$
are a pair of solutions of \eqref{e:eigenequ} with eigenvalue $-\mu^{\ast}$.  In other
words, the eigenvalues come as pairs, $\mu$ and $-\mu^{\ast}$.  Multiplying the bottom
line of \eqref{e:eigenequ} by $-1$ and making use of the Hermitian property of the
operator on the left-hand-side yields an orthogonality relation:
\begin{equation}\label{e:orthogonal}
   ( \mu^{\ast}_i - \mu_j )
   \int \dd[3]{x} 
   [\, a_i^{\ast}(\vb{r}) a_j(\vb{r}) - b_i^{\ast}(\vb{r}) b_j(\vb{r}) \,]
   =
   0 \>.
\end{equation}
Real eigenvalues lead to oscillatory behavior and stability whereas imaginary
eigenvalues lead to blow up or damping and instability of the system. The
system is deemed stable if $\Im{\mu_i} = 0$ for all $i$.  For real eigenvalues,
the states are normalized by:
\begin{equation}\label{e:OrthoNormal}
   \int \dd[3]{x} 
   [\, a_i^{\ast}(\vb{r}) a_j(\vb{r}) - b_i^{\ast}(\vb{r}) b_j(\vb{r}) \,]
   =
   \delta_{i,j} \>. 
\end{equation}
The general solution to \eqref{e:DiffEq} is then given by the sum over all
eigenstates of \eqref{e:eigenequ}:
\begin{align}
   \Phi(\vb{r},t)
   &=
   \begin{pmatrix}
      \phi(\vb{r},t) \\ \phi^{\ast}(\vb{r},t)
   \end{pmatrix}
   =
   \sum_{\text{all $i$}} \Phi_i
      \begin{pmatrix}
         c_i(\vb{r}) \\
         d_i^{\ast}(\vb{r})
      \end{pmatrix} \rme^{-\rmi \mu_i \, t}
   \label{e:GenSol-III} \\
   &=
   \sum_{i > 0} \Phi_i
   \Bigl \{
      \begin{pmatrix}
         a_i(\vb{r}) \\
         b_i(\vb{r})
      \end{pmatrix} \rme^{-\rmi \mu_i \, t}
      +
      \begin{pmatrix}
         b_i^{\ast}(\vb{r}) \\
         a_i^{\ast}(\vb{r})
      \end{pmatrix} \rme^{+\rmi \mu_i^{\ast} \, t} 
   \Bigr \} \>,
   \notag
\end{align}
the last sum now going over the unique eigenvalues only.  At $t=0$,
\begin{equation}\label{e:InitalValue}
   \begin{pmatrix}
      \phi(\vb{r},0) \\ \phi^{\ast}(\vb{r},0)
   \end{pmatrix}
   =
   \sum_{i} \Phi_i
      \begin{pmatrix}
         c_i(\vb{r}) \\
         d_i^{\ast}(\vb{r})
      \end{pmatrix} \>.
\end{equation}
Inverting this relation using \eqref{e:OrthoNormal} 
\begin{align}
   &\int \dd[3]{x}
   \bigl ( \, c_j^{\ast}(\vb{r}), \, d_j(\vb{r}) \, \bigr )
   \begin{pmatrix}
      1 & 0 \\
      0 & -1
   \end{pmatrix}
   \begin{pmatrix}
      \phi(\vb{r},0) \\ \phi^{\ast}(\vb{r},0)
   \end{pmatrix} 
   \label{e:SetIV} \\
   &\hspace{1em}
   =
   \sum_{i} \Phi_i
      \bigl ( \, c_j^{\ast}(\vb{r}), \, d_j(\vb{r}) \, \bigr )
      \begin{pmatrix}
         1 & 0 \\
         0 & -1
      \end{pmatrix}      
      \begin{pmatrix}
         c_i(\vb{r}) \\
         d_i^{\ast}(\vb{r})
      \end{pmatrix}
   =
   \Phi_j \>,
   \notag
\end{align}
gives
\begin{equation}\label{e:SetPhii}
   \Phi_i
   =
   \int \dd[3]{x}
   [\,  c_i^{\ast}(\vb{r}) \phi(\vb{r},0) 
        - d_i(\vb{r}) \phi^{\ast}(\vb{r},0) \,] \>.
\end{equation}
Solutions of the NLSE to first order are then given by
\begin{equation}\label{e:PsiExpand}
   \Psi(\vb{r},t)
   =
   \Psi_0(\vb{r},t) + \varepsilon \, \Phi(\vb{r},t) + \dotsb
\end{equation}
where
\begin{equation}\label{e:Psi0def}
   \Psi_0(\vb{r},t)
   =
   \begin{pmatrix}
      u_0(\vb{r}) \, \rme^{-\rmi \mu_0 t} \\
      u_0(\vb{r}) \, \rme^{+\rmi \mu_0 t}
   \end{pmatrix} \>.
\end{equation}
We must also require that
\begin{equation}\label{e:Psi0Psi1}
   \Psi_0^{\dag}(\vb{r},t) \, M \, \Phi(\vb{r},t)
   =
   0
   \qc
   M
   =
   \begin{pmatrix}
      1 & 0 \\
      0 & -1
   \end{pmatrix} \>,   
\end{equation}
since the unperturbed state is included in $\Psi_0(\vb{r},t)$. This
requirement is usually applied by omitting the $i=0$ state in the
sum appearing in Eq.~\eqref{e:GenSol-III}, as discussed in the next section.

%
%
\subsection{\label{s.LR:oneD}One dimension}

In 1D, the eigenvalue equation \eqref{e:eigenequ} becomes:
\begin{equation}\label{e.LR:eigenequ1D}
   \begin{pmatrix}
     [\, h(x)  + g u_0^2(x) \,] & g u_0^2(x)  \\
     - g u_0^2(x)  & - [\, h(x)  + g u_0^2(x) \,]
   \end{pmatrix}
   \begin{pmatrix}
      a(x) \\
      b(x)
   \end{pmatrix}
   =
   \lambda
   \begin{pmatrix}
      a(x) \\
      b(x)
   \end{pmatrix} \>,
\end{equation}
where $u_0(x)$ is given by Eq.~\eqref{e.2S1D:u0def}, and
\begin{align}
   h(x)
   &= - \partial_x^2 + V_0(x)\>,
   \label{e.LR:hdef-1D} \\
   V_0(x)
   &=
   - a (1 - a q^2) + a^2 x \, (\, x - 2 q \tanh(a q x) \,)\>.
   \label{e.LR:V0def-1D}
\end{align} 
Here we have set $\mu = \mu_0 + \lambda$, and redefined $V_0(x)$
so that now $V_0(0) = \mu_0 = a (1 - a q^2)$. A plot of this potential
as a function of $x$ is shown in Fig.~\ref{f:fig10a} for the case when
$a=1$ and $q=5$. Zero eigenvalues ($\lambda = 0$) now correspond to the
soliton solution $a(x) = -b(x) = u_0(x)$.

%
%
\subsubsection{\label{s.LR:BogApprox}Bogoliubov approximation}

\begin{figure}[t]
   \centering
   \subfigure[\ $V_0(x)$]
   { \label{f:fig10a}
     \includegraphics[width=0.8\linewidth]{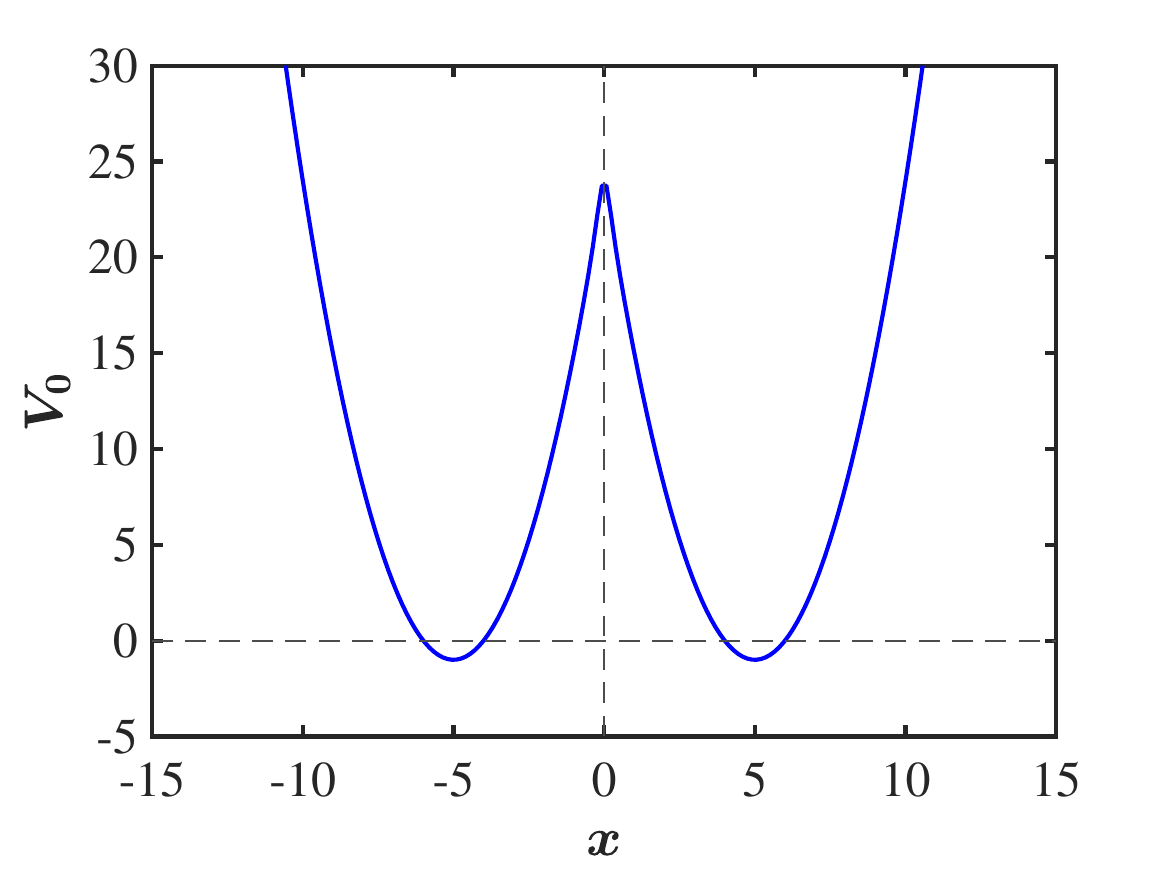} }
   \subfigure[\ $\chi_n(x)$]
   { \label{f:fig10b}
     \includegraphics[width=0.8\linewidth]{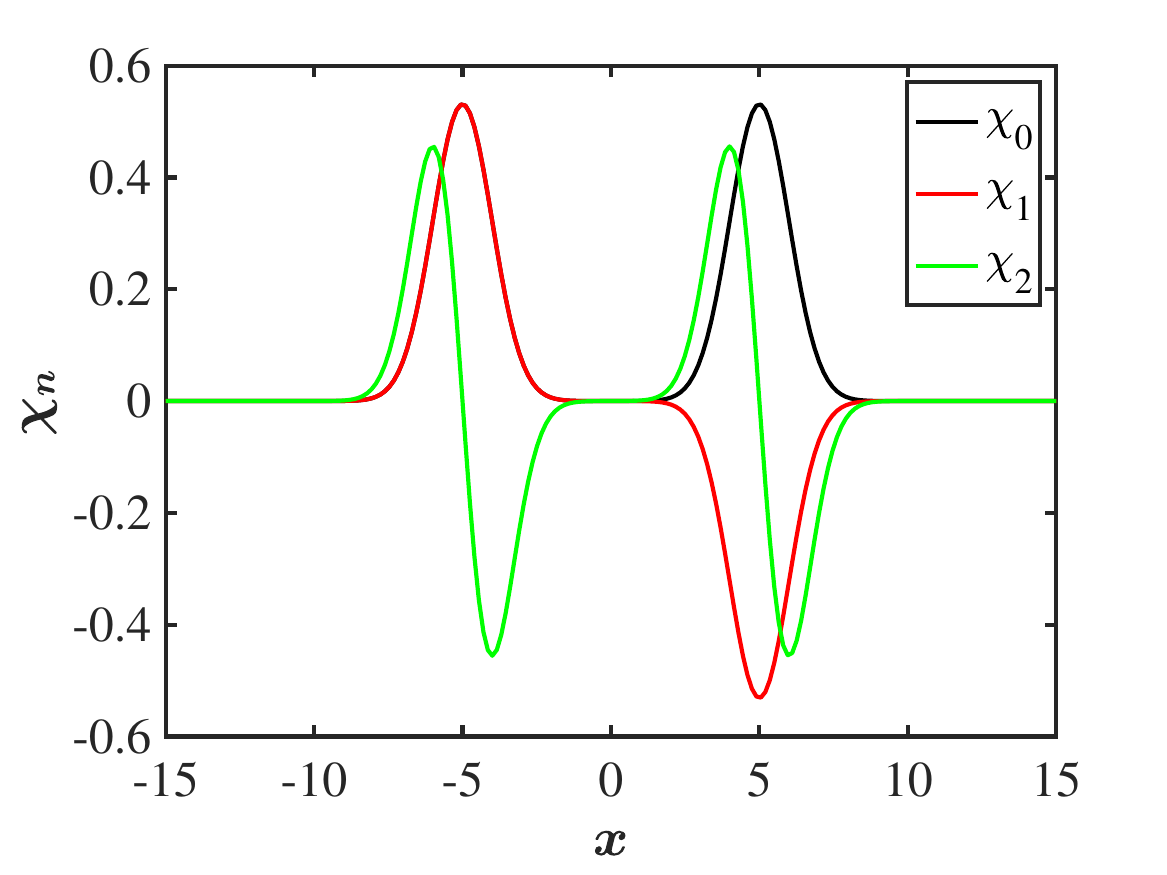} }
   \caption{\label{f:fig10} (a) Plots of the potential $V_0(x)$ in Eq. (62) and
   (b) the wave functions $\chi_n(x)$ in Eq. (64) as functions of $x$ for the
   case when $a=1$ and $q=5$.}
\end{figure}

Moreover, the eigenvalue problem \eqref{e.LR:eigenequ1D} can be
written in terms of eigenvectors of the Hermitian operator $h(x)$ in 1D.
To that effect, we define
\begin{equation}\label{e.LR:heigenprob}
   h(x) \, \chi_n(x) = \epsilon_n \, \chi_n(x) 
   \qc
   \chi_n(x) \in \bbR \>,
\end{equation}
where $\epsilon_n \in \bbR$ and $\chi_n(x)$ obeys the orthonormal relation,
\begin{equation}\label{e.LR:orthonormal}
   \int \dd{x} \chi_n(x) \chi_{n'}(x)
   =
   \begin{cases}
      \delta_{n,n'} \>, & \text{for $n$ and $n'$ $\ne 0$,} \\
      M_0 \>, & \text{for $n = n' = 0$.}
   \end{cases}
\end{equation}
For our case when $a=1$ and $q=5$, the eigenvalues are very close
to being doubly degenerate and are given by $\epsilon_n = 0, 2, 4, 6, \dotsc$
with a small splitting of each state due to tunneling between the two
wells. The low lying spectrum is that of a quantum harmonic oscillator
with frequency $\omega = 2$, as might be expected from the shape of the
double well. A plot of the first few wave functions $\chi_n(x)$ (as functions
of $x$) are shown in Fig.~\ref{f:fig10b}.
%
%
%
%

Expanding the solutions of \eqref{e.LR:eigenequ1D} by setting
\begin{equation}\label{e.LR:abexpand}
   \begin{pmatrix}
      a(x) \\ b(x)
   \end{pmatrix}
   =
   \sum_{n=0}^{\infty} 
   \begin{pmatrix}
      A_n(x) \\ B_n(x)
   \end{pmatrix}
   \chi_n(x) \>,
\end{equation}
and using the orthonormal condition, we obtain the eigenvalue problem
\begin{align}
   &\sum_{n'=0}^{\infty}
   \begin{pmatrix}
      (\epsilon_n - \lambda) \delta_{n,n'} + g \Delta_{n,n'} &
        g \Delta_{n,n'} \\
      - g \Delta_{n,n'} &
      - (\epsilon_n + \lambda) \delta_{n,n'} - g \Delta_{n,n'} 
   \end{pmatrix}
   \notag \\
   & \hspace{3em}
   \times
   \begin{pmatrix}
      A_{n'}(x) \\ B_{n'}(x)
   \end{pmatrix}
   =
   0 \>,
   \label{e.LR:ABeigenvalue}
\end{align}
where
\begin{equation}\label{e.LR:Deltadef}
   \Delta_{n,n'}
   =
   \int \dd{x} u_0^2(x) \, u_n(x) u_{n'}(x) \>.
\end{equation}
The eigenvalues $\lambda$ are then found by solving the determinant:
\begin{equation}\label{e.LR:DetEqu}
   \begin{vmatrix}
      (\epsilon_n - \lambda) \delta_{n,n'} + g \Delta_{n,n'} &
        g \Delta_{n,n'} \\
      - g \Delta_{n,n'} &
      - (\epsilon_n + \lambda) \delta_{n,n'} - g \Delta_{n,n'}       
   \end{vmatrix}
   =
   0 \>.
\end{equation}
Numerical calculations for the case when $a=1$ and $q=5$ gives $\Delta_{1,1} \approx 0.2 \, M_0$ and $\Delta_{2,2} \approx 0.1 \, M_0$, whereas $\Delta_{1,2} = 10^{-9} \, M_0$ and $\Delta_{0,3} = -0.705 \, M_0$.  So a reasonable approximation for the low-lying eigenvalues is to include only diagonal terms, $\Delta_{n,n'} \approx \Delta_{n} \, \delta_{n,n'}$, in which case \eqref{e.LR:DetEqu} becomes
\begin{equation}\label{e.LR:DiagDetEqu}
   \begin{vmatrix}
      \epsilon_n - \lambda + g \Delta_{n} &
        g \Delta_{n} \\
       - g \Delta_{n} &
      - \epsilon_n - \lambda - g \Delta_{n}       
   \end{vmatrix}
   =
   0 \>,
\end{equation}
which gives
\begin{equation}\label{e.LR:DiagResult}
   \lambda_n
   =
   \pm \sqrt{ \epsilon_n ( \epsilon_n + 2 g \Delta_{n} ) } \>,
\end{equation}
which is the Bogoliubov spectrum.  One can see here that for $g=+1$ the system is always stable whereas for $g=-1$ there is a small region of stability as long as
\begin{equation}\label{e.LR:bogStable}
   \epsilon_n \ge 2 g \Delta_{n} \>, 
\end{equation}
for any $n$.  For $n=1$ in our case this means that $2 > 0.4 \, M_0$, or $M_0 < 5$.  For $n=2$, we find $M_0 < 20$, which is a higher bound.  

%
%
\subsubsection{\label{s.LR:Direct}Direct solution of the BdG equation}

The eigenvalue equation of Eq.~\eqref{e.LR:eigenequ1D} is solved numerically in \textsc{MATLAB}
by employing a computational grid in coordinate space, and using a fourth-order
accurate, finite difference approximation for the Laplacian operator. We corroborated
our numerical results by using $P_{3}$ finite elements in the computational
software FreeFEM++\cite{hecht_freefem} that utilizes the ARPACK eigenvalue solver~\cite{arpack},
and obtained similar results.

\begin{figure}[t]
   \centering
   \subfigure[\ $g = +1$]
   { \label{f:fig11a}
     \includegraphics[width=0.9\linewidth]{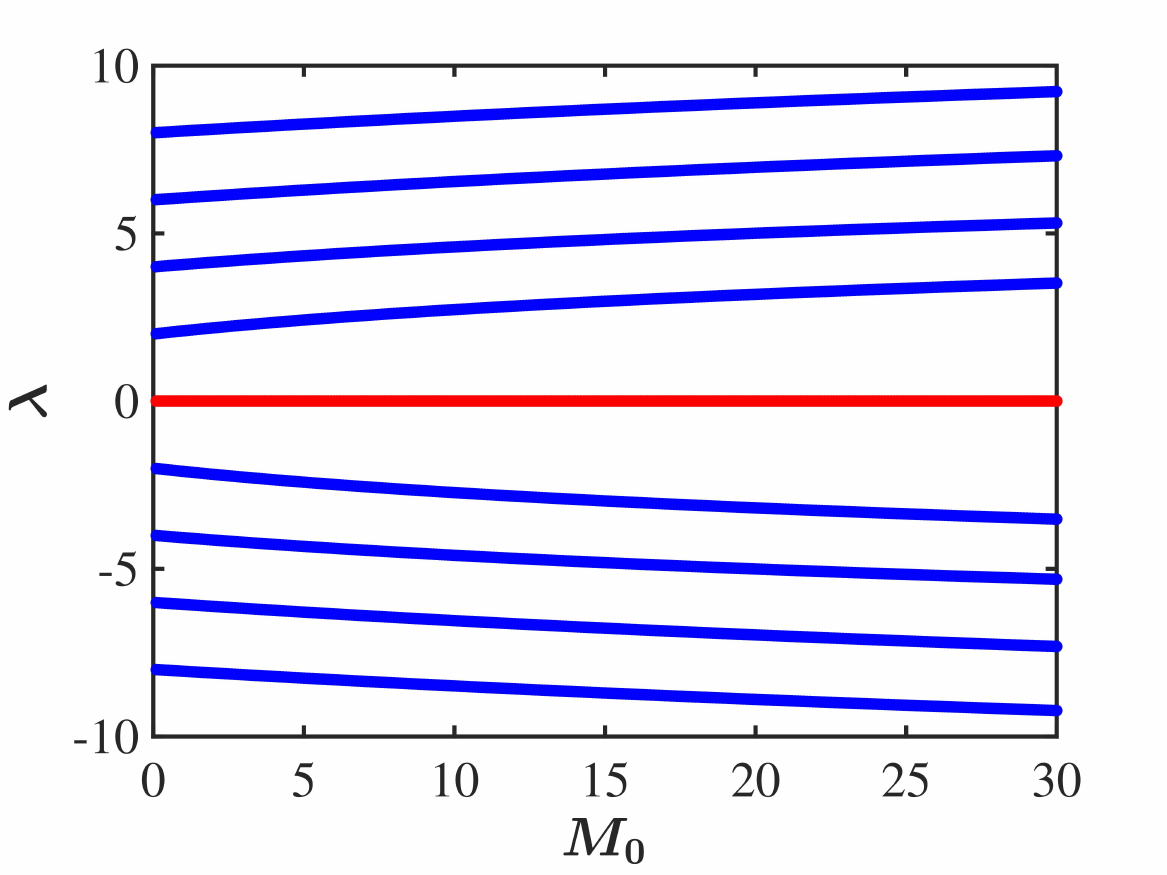} }
   \subfigure[\ $g = -1$]
   { \label{f:fig11b}
     \includegraphics[width=0.9\linewidth]{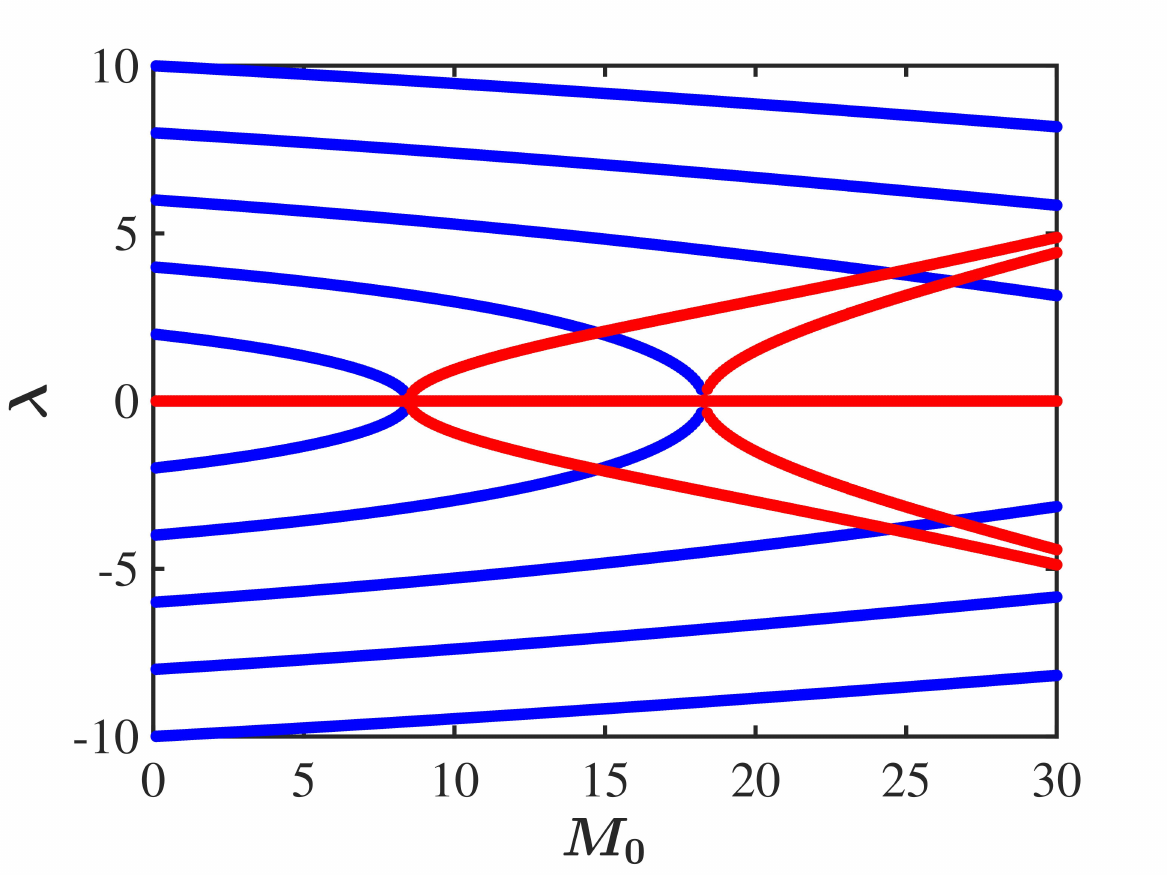} }
   \caption{\label{f:fig11}Real (blue lines) and imaginary (red lines) 
   parts of the eigenvalues $\lambda$ in Eq.~\eqref{e.LR:eigenequ1D} as functions
   of $M_0$ for the one-dimensional case with $a=1$ and $q=5$.}
\end{figure}

\begin{figure}[t]
   \centering
   \includegraphics[width=0.9\linewidth]{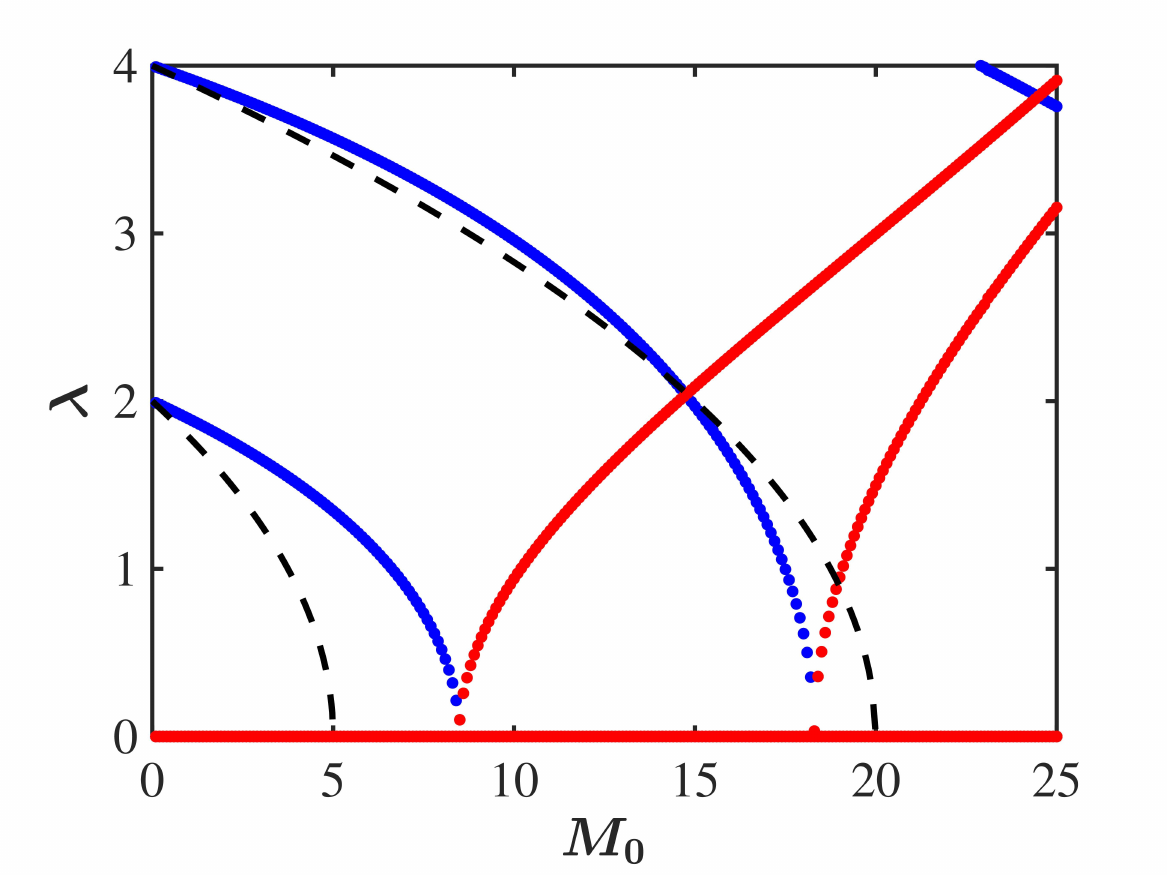}
   \caption{\label{f:fig12}Comparison of the real part of the $g=-1$
   numerically (exact) eigenvalues $\lambda$ (in blue) with the Bogoliubov
   formula (in dashed black) of Eq.~\eqref{e.LR:DiagResult}.}
\end{figure}

In 1D, numerical results for the eigenvalues $\lambda$ of this calculation are plotted
in Fig.~\ref{f:fig11} as functions of $M_0$ for $g=\pm 1$ and parameter values $a=1$ and
$q=5$. The real part of the eigenvalues is shown in red whereas their imaginary part is
shown in blue. The top panel of the figure corresponding to the repulsive ($g=1$) case
suggests that the solution is linearly stable. On the other hand, when we consider attractive
interactions, i.e., $g=-1$, the solution is (linearly) stable but becomes unstable past
$M_0 \gtrsim 8$, in approximate agreement with the Bogoliubov approximation and Derrick's
theorem (see Section~\ref{ss:1D}). At $M_0 = 0$, the low-lying eigenvalues are all real
and given by $\lambda_n \approx 0, 2, 4, 6,\dotsc$.
%
%
%
%
Moreover, we compare the (1D) numerical results of Eq.~\eqref{e.LR:eigenequ1D}
for $g=-1$ (see, the bottom panel of Fig.~\ref{f:fig11}) with the approximate
Bogoliubov result from Eq.~\eqref{e.LR:DiagResult} in Fig. \ref{f:fig12}.
The shape of the Bogoliubov curve shown with dashed black line is
proximal to the numerically computed eigenvalues although the point
in the parameter space where the solution is predicted to be unstable
is at lower values of $M_{0}$. This is somewhat expected because not
enough terms were included in Eq.~\eqref{e.LR:DetEqu} which itself
would allow otherwise a better agreement between the two approaches.
%
%
%
%

%
%
\begin{figure}[t]
   \centering
   \subfigure[\ $g = +1$]
   { \label{f:fig13a}
   \includegraphics[width=0.9\linewidth]{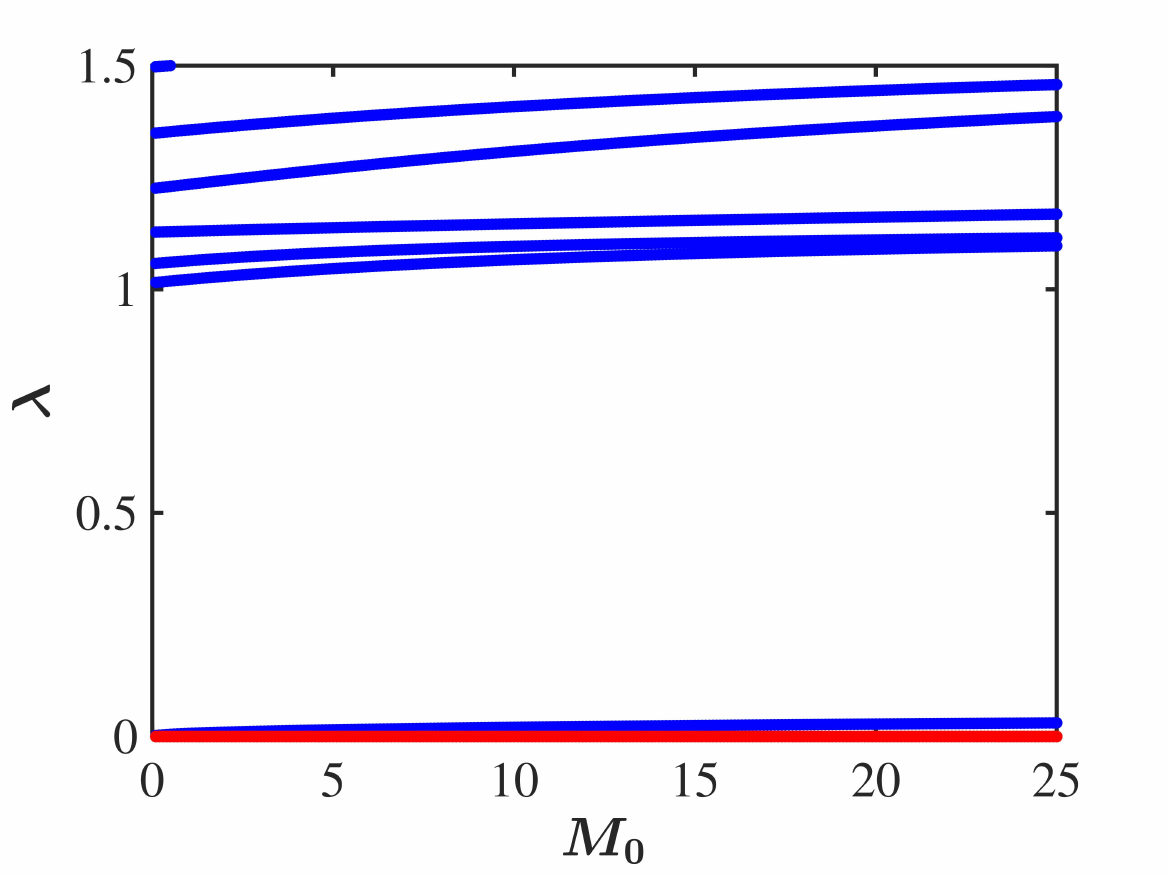} }
   %
   \subfigure[\ $g = -1$]
   { \label{f:fig13b}   
   \includegraphics[width=0.9\linewidth]{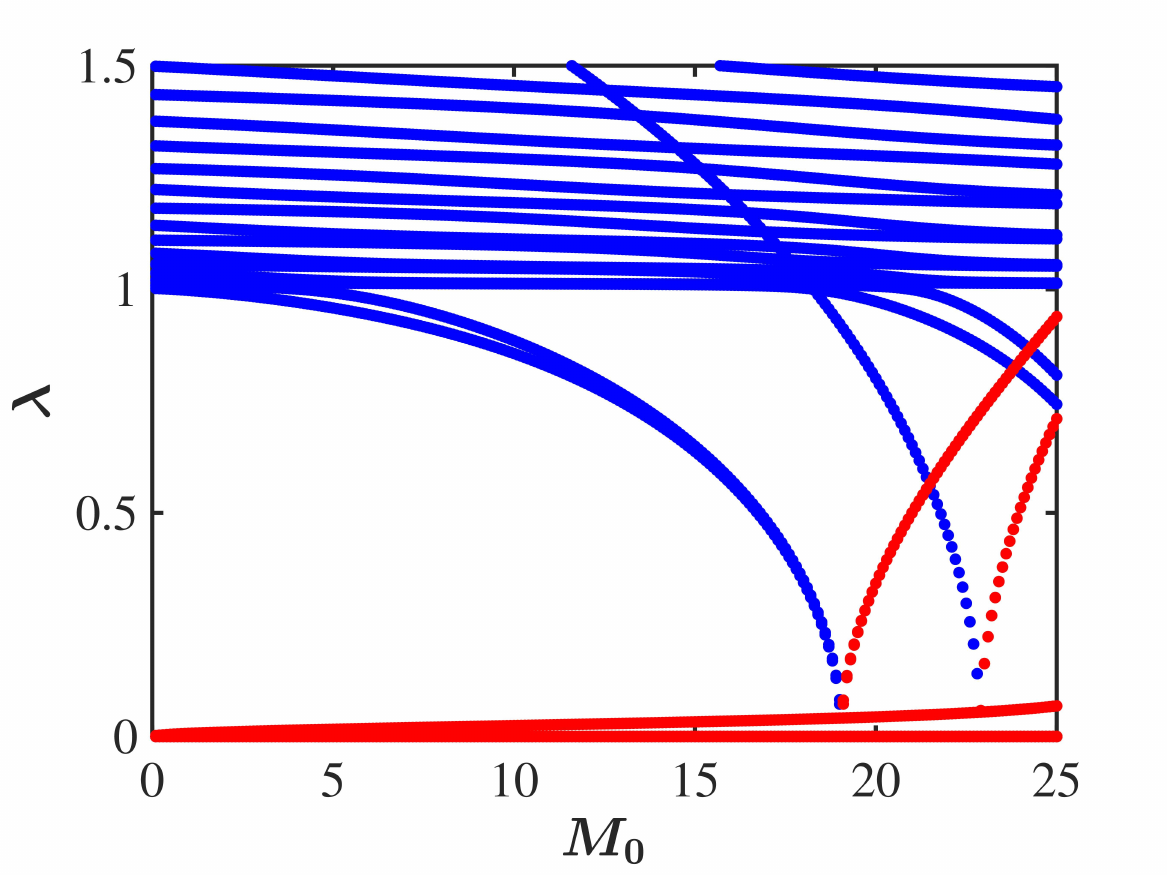} }
   \caption{\label{f:fig13}The real part (in blue) and imaginary part
   (in red) of the eigenvalues for the two-dimensional soliton of Eq.~\eqref{e.2D:u02}
   as functions of $M_{0}$ for the $g=\pm 1$ cases, with $a=b=1$ and $q=5$.}
\end{figure}
%
%

Similar conclusions are drawn in the 2D case (see, Sec.~\ref{ss:2D}).
To that end, we briefly discuss our findings in Fig.~\ref{f:fig13}
which depicts numerical results for the eigenvalues $\lambda$
(see, also Sec.~\ref{ss:2D}) when $a=b=1$ and $q=5$. Again the
system is stable when $g=-1$ for all values of $M_0$, whereas
when $g=+1$ there is a region of stability for $M_0 \lesssim 7.5$.
We note in passing that we have checked the stability and instability
(over the respective parameter regime) of the solutions that we have
presented so far by performing direct numerical simulations of the
GPE [cf. Eq.~\eqref{e:NLSE}] although we omit the presentation of
the respective results herein.

%
%
\section{\label{s:conclusions}Conclusions}

In this paper, we considered the nonlinear \Schrodinger\ equation (NLSE) or Gross-Pitaevskii
equation (GPE), and employed the ``inverse problem'' method for the potential therein. We
discussed various such external potentials in 1D and higher spatial dimensions, and obtained
respective exact yet confined solutions. We showed that these solutions to the NLSE may possess
an arbitrary number of ``soliton-like'' maxima, i.e., $N$-soliton solutions. Since these
solutions are exact, we can obtain analytical estimates for the values of the $L^2$ norm of
the solution or, equivalently, the number of atoms in the trap above which these solutions are
unstable to width perturbations using Derrick's theorem. We further solved numerically the underlying
eigenvalue (BdG) problem emanating from the linearization of the NLSE whose results are in line
(in terms of stability characteristics) with the theoretical predictions from Derrick's theorem.
However, in all cases that we have studied in this work that turned out to be unstable (i.e.,
attractive interactions), the BdG analysis gives a lower value for the critical value of the norm of the wave function than Derrick's theorem or the translational instability. To the extent that we can identify these distinct entities as separate BEC solutions of the Gross-Pitaevskii equation, then we have given a prescription
for what external potentials will produce various configurations of BECs that are stable in
1D and higher spatial dimensions.

%
%
%
\appendix 
%
%
\section{\label{s:OtherSolutions}Some other $N$-soliton solutions in 2D}

\begin{figure} [t]
 \centering
   \subfigure[Density]
   {\label{fig:rho3sol}
	\includegraphics[width=0.74\linewidth]{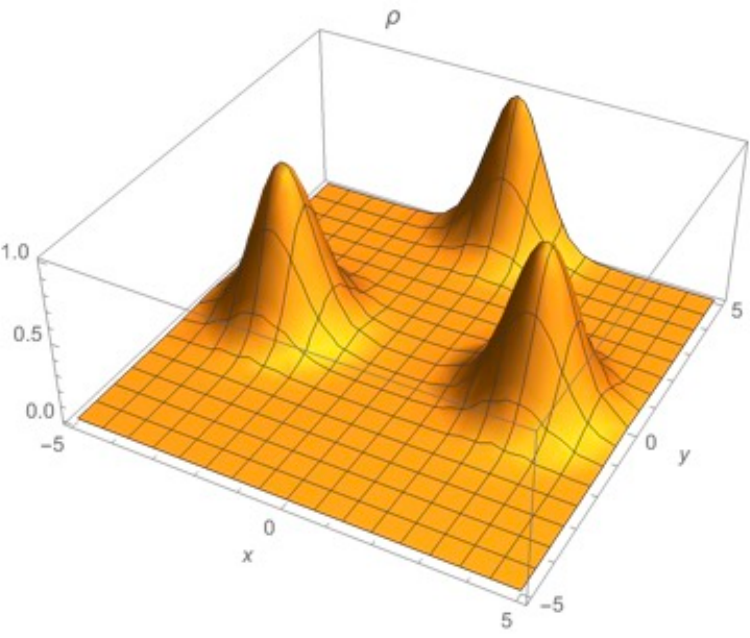}}
   \hspace{0.1em}
   \subfigure[Confining potential]
	{\label{fig:V3sol}
	\includegraphics[width=0.74\linewidth]{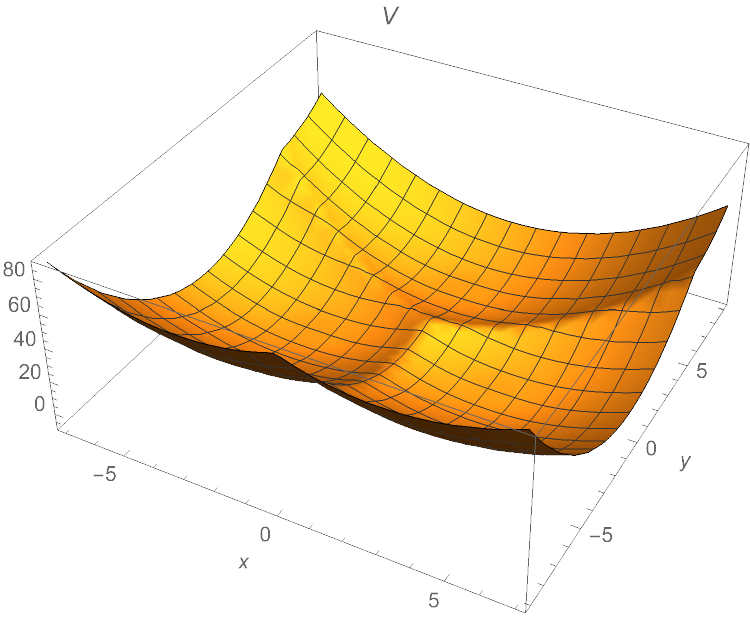}}
   \caption{\label{f:rhoV3sol} The density $\rho(x,y)$  and the confining 
   potential $V(x,y)$ both as functions of $x$ and $y$ for the
   three soliton case, when $g=-1, a=1,q=3$, and $A=1$.}
\end{figure}

\begin{figure} [t]
 \centering
   \subfigure[Density]
	{\label{rho9sol}\
	\includegraphics[width=0.74\linewidth]{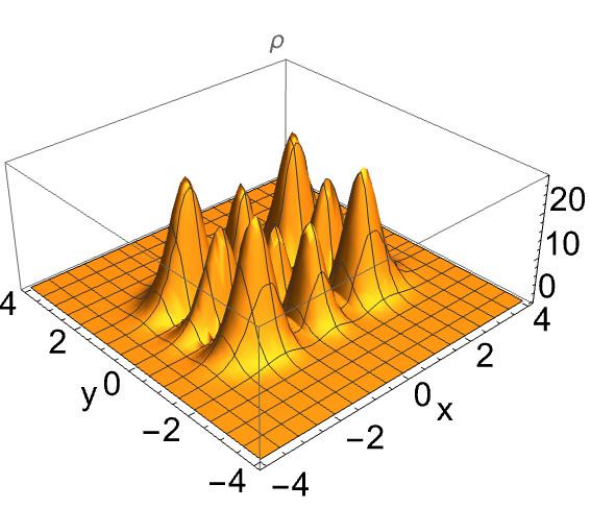}}
   \hspace{0.1em}
   \subfigure[Confining potential]
	{\label{V9line}
	\includegraphics[width=0.74\linewidth]{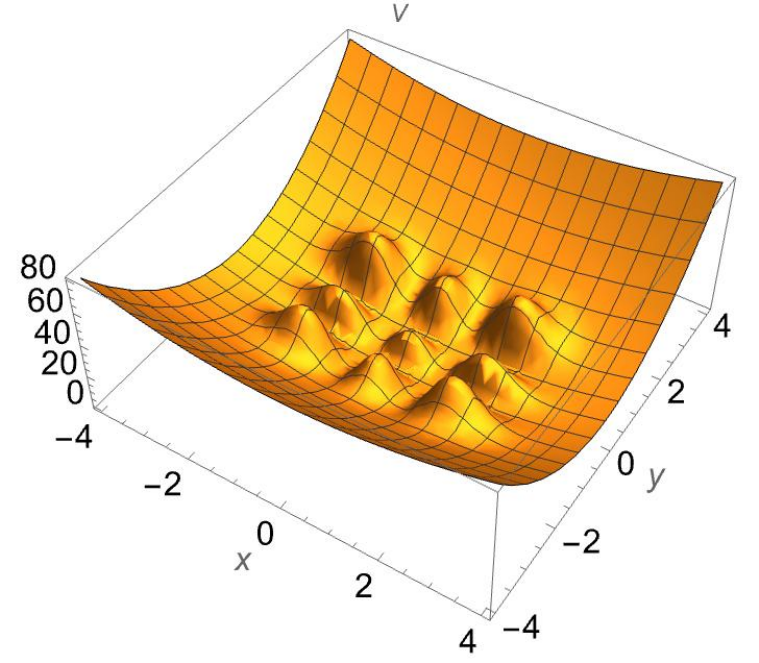}}
   \caption{\label{f:rhoV9sol} The density $\rho(x,y)$ and the confining 
   potential $V(x,y)$ both as functions of $x$ and $y$, for the nine
   soliton case, when  $m=n=2$ and
   $g=-1,A=1,a=1$, and $b=2$.}
\end{figure}

It is clear that there are infinite possibilities for exact $N$-soliton
solutions in 2D and 3D. Here we will give two examples not included in the main
text. For the sum of Gaussians, it is easy to generalize to $N$ soliton exact
solutions. As an example of this, consider the case where the solitons are centered
at the ends of an equilateral triangle. That is, we take for the initial condition:
\begin{align}
   u(x,y) 
   &=
   A
   \Bigl \{\, 
      \rme^{- a (y-\sqrt{3} q)^2/2 - a x^2/2}
      \\
      & \hspace{2em}
      +
      \rme^{ - a y^2/2} \,
      \Bigl [\,
         e^{ - a (x-q)^2/2}
         +
        e^{ - a (q+x)^2/2} \,
      \Bigr ]\, 
   \Bigr \} \>,
   \notag
\end{align}
and obtain from the inverse method:
\begin{equation}
   \mu_0
   = 
   2 a - a^2 q^2 - \frac{a^2 q^2}{e^{a q^2} + 1/2} \>,
\end{equation}
together with $V(x,y) = V_1(x,y) + V_2(x,y)$ where
\begin{align}
   V_1(x,y)
   &=
   - A^2 g \rme^{-a (3 q^2  + 2 q x + x^2 + y^2)}
   \\
   & \hspace{1em}
   \times
   \Bigl [\,
      \rme^{a q^2}+e^{a q (x+\sqrt{3} y )}
      +
      \rme^{a q (q+2 x)} \,
   \Bigr ]^2 
   \notag
\end{align}
and
\begin{align}
   V_2(x,y) 
   &= 
   \frac{a^2}
      {\bigl ( 2 e^{a q^2}+1 \bigr ) 
       \bigl (
          \rme^{a q^2} + \rme^{a q (x+\sqrt{3} y )}
          +
          \rme^{a q (q+2 x)}
       \bigr )} 
   \notag \\
   & \hspace{-2em} \times
   \Bigl \{\,
      \rme^{a q (q+2 x)} 
      ( -2 q^2 - 2 q x + x^2 + y^2 )
      \notag \\
      & \hspace{1em}
      +
      2 \rme^{ 2 a q^2 }(2 q x + x^2 + y^2)
   \\
   & \hspace{1em}
   + 
   \rme^{a q^2} (-2 q^2+2 q x+x^2+y^2 )
   \notag \\
   & \hspace{1em}
   +
   2 (2 q^2-2 \sqrt{3} q y+x^2+y^2) \, \rme^{a q (q+x+\sqrt{3} y)}
   \notag \\
    & \hspace{1em}
    +
    2 \rme^{2 a q (q+x)} ( -2 q x+x^2+y^2 ) 
   \notag \\
    & \hspace{1em}
    + 
    (-2 \sqrt{3} q y+x^2+y^2) e^{a q (x+\sqrt{3} y)}
   \Bigr \} \>.
   \notag
\end{align}
For the choice of parameters values $g=-1,A=1,a=1$, and $q=3$,
we depict the density $\rho(x,y)$ and potential $V(x,y)$ in Fig.~\ref{f:rhoV3sol}.
One can also have solitons along both the $x$- and $y$-axes
by choosing
\begin{equation}
   u_0(x,y) = A H_n(\sqrt{a}x) \, H_m(\sqrt{b} y) \, 
   \rme^{-a x^2/2 - b y^2/2} \>,
\end{equation}
for which one has $n+1$ solitons in $x$-direction and $m+1$ solitons in
$y$-direction where
\begin{align}
   V(x,y) 
   &= 
   a^2 x^2 + b^2 y^2 
   \\
   & \hspace{2em}
   - 
   g A^2 \, H_{n}^{2}(\sqrt{a} x) \, H_{m}^{2}(\sqrt{b}y) \,
   \rme^{-ax^2/2 -by^2/2} \>,
   \notag
\end{align}
with $\mu_0 = (2n+1) a+(2m+1) b$.
When $m=n=2$, the trapped solution has 9 peaks. For that case we obtain
\begin{subequations}\label{e:ninepeaks}
\begin{align}
   M_0
   &=
   64 \pi  A^2/\sqrt{a b} \>,
   \\
   \rho(x,y)
   &=  
   A^2 (2-4 a x^2)^2 \, (2-4 b y^2)^2 e^{-a x^2-b y^2}\>, 
   \\
   V(x,y) 
   &= 
   a^2 x^2 + b^2 y^2
   \\
   & \hspace{2em}
   - 
   16 \, A^2  g (1-2 a x^2)^2 \, (1-2 b y^2)^2 e^{-a x^2-b y^2} \>,
   \notag
\end{align}
\end{subequations}
and with $\mu_0 = 5 (a+b)$.
An example of this for $m=n=2$ and $A=1,a=1,b=2$, and with $g=-1$ is shown in Figs~\ref{f:rhoV9sol}.
Derrick's theorem in this case allows us to determine the critical mass
for instability, which is given by:
\begin{equation}
   M_c = \frac{81920 \, \pi  (a+b)}{11029 \sqrt{a b}}\>.
\end{equation}
For $a=1, b=2$, and $g=-1$ we find:
\begin{equation}
   M_c = \frac{122880 \, \sqrt{2} \pi }{11029} \approx 49.5.
\end{equation}
%

%
%
\acknowledgments
The motivation for working on this problem resulted from discussions that one of us
(FC) had with Alan Chodos on using BECs to test modifications of gravity. FC, JFD,
and EGC would like to thank the Santa Fe Institute and the Center for Nonlinear
Studies at Los Alamos National Laboratory for their kind hospitality. One of us
(AK) is grateful to Indian National Science Academy (INSA) for the award of INSA Honorary
Scientist position at Savitribai Phule Pune University. The work of EGC has been
supported by the U.S. National Science Foundation under Grants No. DMS-2204782.
The work at Los Alamos National Laboratory was carried out under the auspices of
the U.S.~DOE and NNSA under Contract No.~DEAC52-06NA25396.

%
%
\bibliography{Chaos-11.bib}
%
%
\end{document}